# Embossing of Tapered, Isometric Micro-Interdigitated Flow Fields for Enhanced Desalination by Faradaic Deionization

Md Habibur Rahman,[a] Sadia Afrin,[a] and Kyle C. Smith[a,b,c,*]

mhr7@illinois.edu, safrin2@illinois.edu, and kcsmith@illinois.edu

ORCID IDs: https://orcid.org/0000-0003-1936-1917, https://orcid.org/0009-0008-6031-4171, and https://orcid.org/0000-0002-1141-1679

[a]Department of Mechanical Science and Engineering, Grainger College of Engineering, University of Illinois at Urbana-Champaign, Urbana, IL 61801

[b]Department of Materials Science and Engineering, Grainger College of Engineering, University of Illinois at Urbana-Champaign, Urbana, IL 61801

[c]Computational Science and Engineering Program, Grainger College of Engineering, University of Illinois at Urbana-Champaign, Urbana, IL 61801

*Corresponding author's contact information:

Name: Professor Kyle C. Smith

Address: 1206 W. Green St., Urbana, IL 61801

Phone: +1-217-300-0928

Email: kcsmith@illinois.edu

**Table of Contents**

**Table of Contents Graphic**

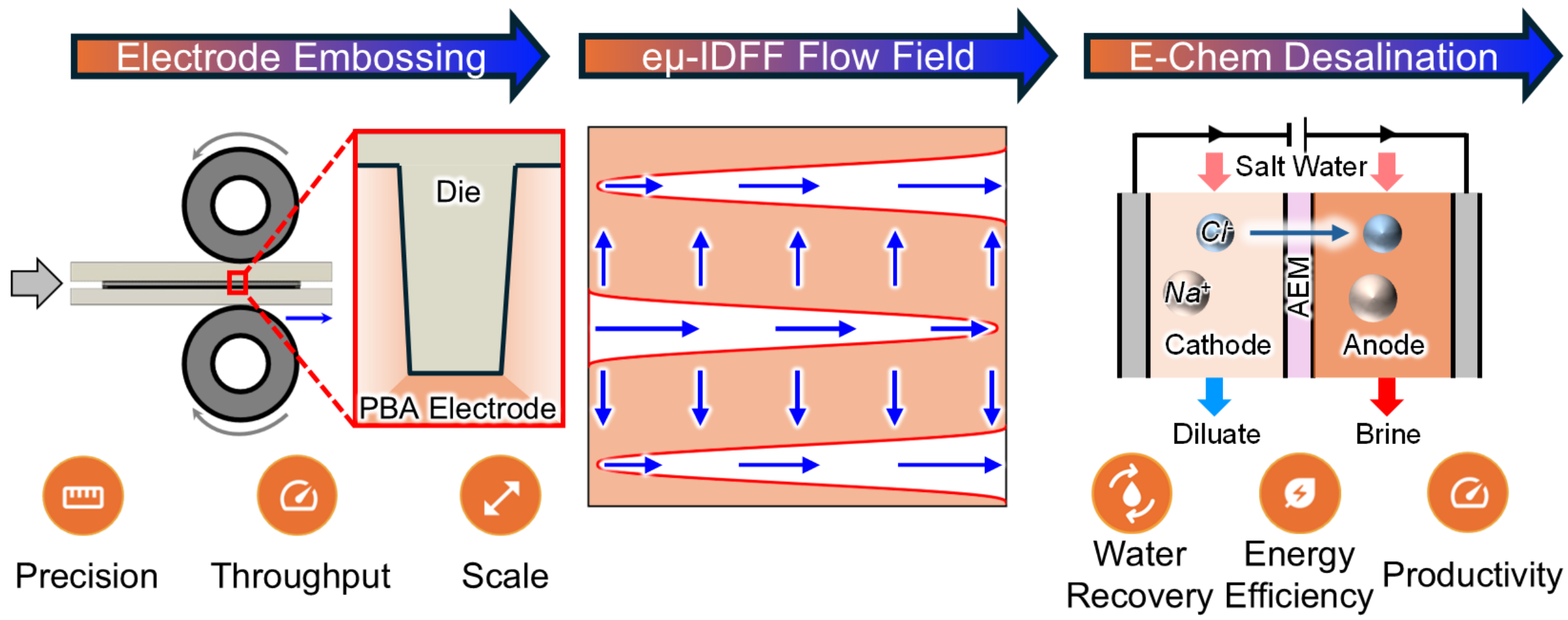

Embossing is used to pattern porous electrodes with tapered, isometric micro-interdigitated flow fields with roll-to-roll compatibility. Patterning is 100-fold faster than milling and needs 15-fold less force than stamping, while preserving active material. The resulting embossed electrodes distribute flow uniformly and desalinate seawater-salinity feeds to near-potable levels with 1.5-fold higher thermodynamic energy efficiency and water recovery than milled, non-isometric designs.

**Abstract**

Engineered transport architectures can improve electrolyte distribution and active-material utilization in porous electrodes, yet their implementation remains limited by fabrication methods that compromise precision, throughput, active-material preservation, or scalability. Here, an integrated embossing-calendering method is introduced to pattern Prussian blue analog (PBA) porous electrodes with embedded, micro-interdigitated flow fields (eμ-IDFFs) for electrochemical desalination. This non-subtractive, roll-to-roll compatible method forms channels by controlling plastic deformation while simultaneously calendering the electrode, preserving active material and reducing patterning time by 100-fold relative to mechanical milling. Indentation experiments and a closed-form model further show that embossing requires 15-fold less force than stamping, while electrode microstructures show local densification beneath embossed channels. PBA electrodes are embossed with novel eμ-IDFFs that use continuous width-tapering with isometric flow paths that promote uniform electrolyte distribution, increase rate capability, and reduce polarization. Compared with eμ-IDFFs patterned using less scalable means and non-isometric flow paths, these eμ-IDFFs desalinate seawater-salinity feeds near to potable levels with 1.5-fold of its thermodynamic energy efficiency and water recovery. Brackish desalination benefits similarly. Tapered, isometric eμ-IDFFs and the embossing method introduced to pattern them are readily translated other electrochemical energy storage and separations systems.

## 1. Introduction

Electrochemical devices that rely on flow through porous electrodes, including fuel cells,[1] redox flow batteries,[2,3] and electrochemical desalination systems,[4,5] are increasingly limited not only by active material chemistry but also by electrode structure. Traditionally, porous electrodes have been designed to maximize electrochemically active surface area, while maintaining sufficient electrolyte accessibility and electronic conductivity.[6,7] More recently, however, insights from physics-based modeling have been used to tailor electrode porosity to improve electrolyte distribution, minimize stagnant regions (i.e., dead zones),[8,9] reduce transport resistance,[10,11] mitigate concentration gradients,[12] and increase active-material utilization.[13] These findings have shifted attention toward engineering multi-dimensional transport pathways. However, translating model-informed designs into practical systems remains a challenge because existing electrode structuring methods often struggle to fabricate complex architectures with the precision and scalability needed for practical implementation.[5,14]

To address transport limitations within porous electrodes, numerous structuring methods have been explored to create engineered pathways for efficient electrolyte and ion distribution.[15] Early efforts focused primarily on reducing electrode tortuosity in lithium-ion batteries through the incorporation of straight channels or pores using subtractive methods such sacrificial templates[16–22] or laser machining.[23–25] Subsequent studies expanded these concepts through additive manufacturing (direct-ink-writing),[26–28] mechanical templating,[29–34] and acoustophoretic assembly[35,36] that provided increasing geometric flexibility. Similar patterning concepts have also been explored in ion-exchange membranes to enhance mass transport and reduce hydraulic or ionic resistance in electrodialysis[37–41] and membrane capacitive deionization.[42] Collectively, these studies have demonstrated that engineered transport pathways can significantly improve electrochemical performance across a range of systems.

Despite this progress, existing fabrication methods remain constrained by certain trade-offs. Subtractive methods such as laser machining and mechanical milling inevitably remove electrochemically active material[14,15,43,44] while potentially introducing thermal or mechanical damage. Laser-perforated electrodes, which are limited to ~200 μm features based on a given beam width,[23] can introduce heat-affected zones due to diffuse laser intensity profiles[5] that melt or otherwise damage active material.[24] In turn, depth-wise heat penetration can reduce adhesion to current collectors.[25] Mechanically milled trenches in capacitive electrodes were also shown to reduce tortuosity by 40-60% at the expense of 20-30% loss in electrode mass.[14] Further, the trench depth achieved by milling is jointly determined by electrode flatness and tool durability, while requiring ~10 hours to mill a single 20 $cm^2$ electrode.[45] Sacrificial-template methods avoid material loss by introducing a temporary channel-forming phase that is later removed by pyrolysis,[16,17] dissolution,[18] or phase change,[19,20] but these methods often require refractory electrode constituents. Additive manufacturing promises design freedom in principle but is limited in practice by feature resolution that is set by nozzle diameter,[26,28] specialized ink rheology,[27,46] post-print coagulation,[46,47] and high-temperature steps[27] that limit throughput and constrain electrode-material selection. Mechanical templating methods preserve active material while offering orders higher throughput and resolution below 100 μm,[32,34] but demonstrations to date have required substantial applied pressure[48] or have largely been restricted to lab-scale processing.[30,31] Figure 1a shows that electrode patterning and structuring methods that simultaneously combine precision, active-material preservation, high throughput, and scalability are non-existent today.

In parallel with the development of methods for electrode patterning, improved fluid distribution has been sought in electrochemical flow devices using increasingly complex flow-field designs. Conventional parallel[49,50] and serpentine[9] flow-fields have been supplemented by interdigitated,[10,51] hierarchical,[52] topology-optimized,[53] machine-learning assisted,[50,54] artificial-intelligence assisted,[55] biomimetic,[56–60] and lung-inspired[61,62] flow-field designs that seek to improve reactant distribution while reducing hydraulic losses. Many such studies have demonstrated improvements in electrochemical performance (e.g., reduced voltage polarization) through improved flow distribution and mass transport, but these gains often come at the expense of increased fabrication time and manufacturing cost that are not practical beyond prototyping. Metal-sintered, lung-inspired, and biomimetic plates, for example, cost ~$1k and require up to ~1 week to fabricate.[56,62] These examples highlight a disconnect between advances in flow/transport architecture and the patterning/structuring methods available to realize

them at scale. Bridging this gap requires methods that accurately reproduce complex, multi-scale designs while preserving electrochemically active material, integrating with established electrode-processing workflows, and remaining compatible with scalable production. Developing such methods would facilitate the practical implementation of patterned and structured electrodes across a range of electrochemical technologies.

We presently introduce an integrated embossing-calendering method for the first time to pattern porous electrodes that contain embedded, micro-interdigitated flow fields (eμ-IDFFs).[5,45] The precision and throughput provided by embossing are exploited to pattern electrodes with novel eμ-IDFFs that include 4.5 cm-long channels whose width is continuously tapered down to ~10 μm to effect uniform electrolyte flow along 640 μm flow paths through a fixed 200 μm depth. This roll-to-roll compatible method is shown to form flow channels using controlled plastic deformation while simultaneously densifying electrode material during calendering and preserving active material. This method is used to embed eμ-IDFFs into electrodes containing nickel hexacyanoferrate Prussian blue analog (PBA) nanoparticles that drive water desalination using a symmetric Faradaic deionization (FDI) process. Here, the associated tapered channels that comprise a given eμ-IDFF are designed to uniformize saltwater flow through electrodes, while mitigating limitations of earlier tapered-channel designs that produced variable flow-path length.[5,45] Compared to mechanical milling, the proposed process reduces fabrication time greater than 100-fold. Finally, the salt removal, energy consumption, and production rates obtained using the associated embossed PBA electrodes to desalinate water are shown to improve substantially on past work that used less scalable patterning methods.

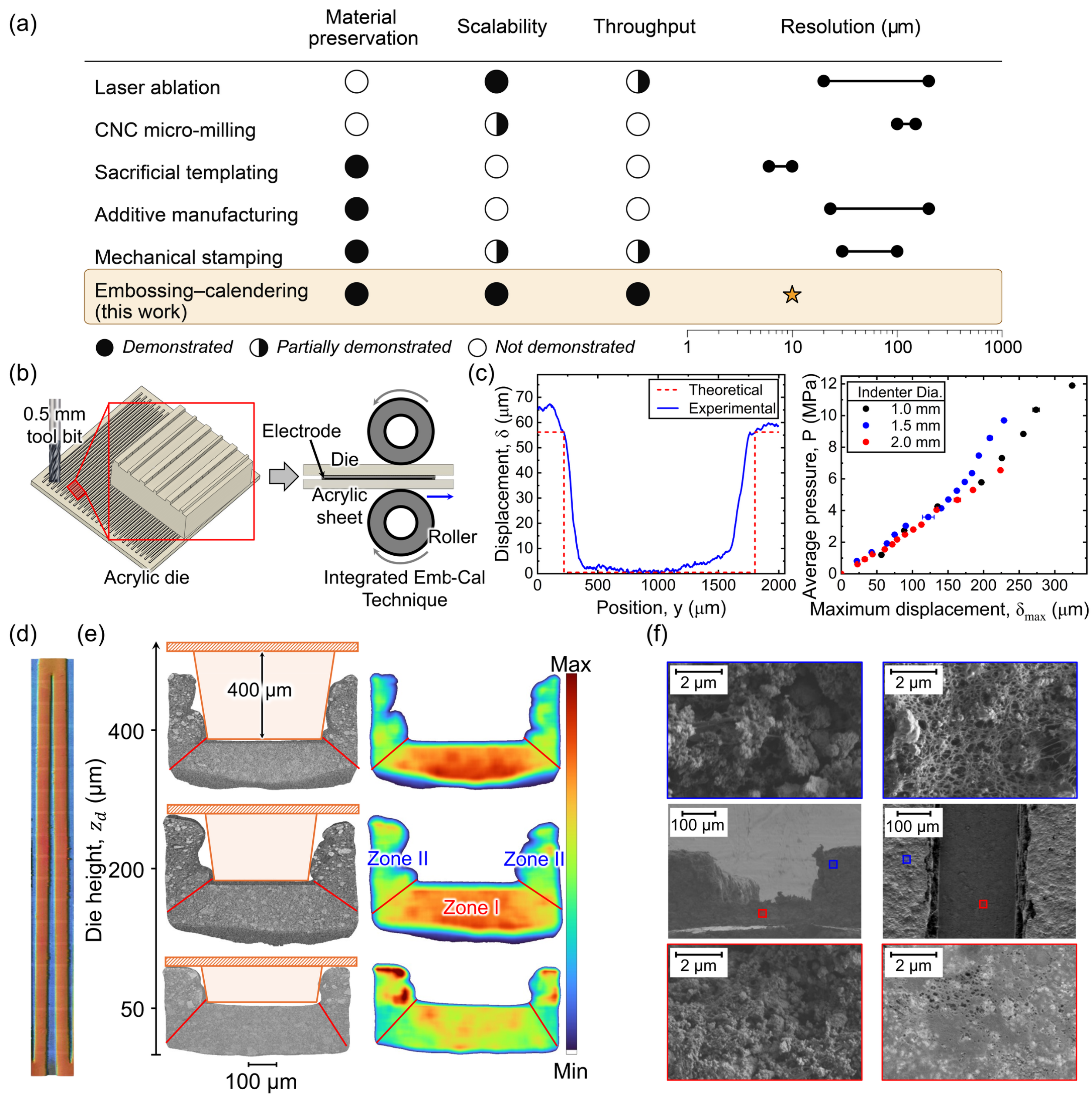

**Figure 1:** (a) Electrode patterning methods compared using four criteria: material preservation, scalability, throughput, and resolution. (b) Integrated embossing-calendering process using a die with the associated negative shape to pattern eμ-IDFFs into porous electrodes. (c) Electrode height versus lateral position for indentation with a 1.5 mm indenter whose profile (left) and average pressure versus maximum displacement for indentation using certain indenter diameter (right). (d) Height map of embossed Gen. III interdigitated channels having a continuous, isometric width profile. (e) Raw and box-filtered X-ray computed tomography (CT) cross-sections for electrodes embossed using die heights of 50, 200, and 400 μm. (f) Microstructure of electrodes embossed to 200 μm depth, as determined from scanning electrode microscopy (SEM). Red boxes show views beneath the embossed channel (bottom-left) and along the channel's bottom surface

(bottom-right), whereas blue boxes show views near the channel's edge (top-left) and at the top electrode surface (top-right).

## 2. Results and Discussion

### *2.1. Mechanics and Microstructure of Electrode Embossing*

We first introduce and characterize a scalable, non-subtractive embossing method to pattern eμ-IDFFs onto porous electrodes. In contrast with stamping that forms all features simultaneously under dynamic uniaxial compression, embossing leverages tangential friction in concert with compression as a given electrode is drawn between two calender rolls (Fig. 1b). In an up-scaled embodiment of embossing the associated rollers may be engraved with the negative shape of the eμ-IDFFs of interest to produce a continuous roll-to-roll process, like in gravure printing.[63] Here, we instead employ a separate mechanically milled die including the associated negative pattern that enters with an electrode between the rollers to enable the prototyping of various eμ-IDFF generations that are described in Sec. 2.2 (Fig. 1b).

To inform implementation of this non-subtractive patterning method we analyzed the mechanics and microstructure of Prussian blue analog (PBA) porous electrodes subjected to uniaxial cylindrical indentation using varied indenter size and *ex situ* optical profilometry. Figure 1c,left shows that indentation produces a height map that closely matches a given indenter's profile. While such deviations indicate a degree of elastic recovery after indentation, their small magnitude relative to the indenter's diameter indicates that the indenter's forces are primarily accommodated by permanent, plastic deformation. This observation is consistent with previous grid indentation experiments on lithium-ion battery electrodes that showed 10-fold smaller hardness than elastic modulus.[64] Embossed channels also possess width and depth deviations smaller than 30 μm from their intended cross-sections (Fig. S2). The average applied pressure, calculated as the normal force imposed during indentation divided by the indenter's tip area, is also shown to scale linearly with permanent surface displacement across all tip sizes (Fig. 1c,right and Fig. S3), indicating that plastic compression occurring beneath the indenter produces larger forces than plastic shear at the indenter's edges.

By assuming linear scaling between pressure and displacement with a proportionality constant $\kappa$ equal to 36 GPa/m, we also developed a closed-form mechanical model in SI Sec. 6 to predict the force required to pattern eμ-IDFFs onto porous electrodes using either embossing or stamping. The compression force $F_R' = (2/3)\kappa\Delta h\sqrt{2R\Delta h}$ per unit width required to emboss an electrode using rolls of radius $R$ follows a three-halves power law in the thickness change $\Delta h$ that embossing produces, while the corresponding torque $T_R' = r_{eff}F_R'$ per unit width is determined using the effective moment arm $r_{eff} = (3/8)\sqrt{2R\Delta h}$. In contrast, the compression force $F_S' = \kappa L\Delta h$ per unit width required to stamp an electrode is found to scale linearly with $\Delta h$. The ratio of stamping-to-embossing force, $F_S'/F_R' = (3/2)\sqrt{L^2/(2R\Delta h)}$, is 18.9 for the rolls of 3.2 cm radius used presently with a nominal 200 μm thickness change on a 45 mm-long electrode. This force ratio is increasingly shown to favor embossing for longer electrodes and smaller rolls. This model also shows that the stamping force required to pattern the tapered, isometric eμ-IDFF introduced later is 15-fold larger than for embossing (Fig. S7).

We also used X-ray computed tomography (CT) to characterize the microstructure of porous electrodes embossed with dies of fixed maximal width (514 μm) and three heights (50, 200, 400 μm), with roller gap tuned for each die height. In all cases the cross-sections of fine-resolution CT intensity (Fig. 1e) show high-contrast PBA aggregates with <10 μm size in a low-contrast background constituted by conductive additive and binder. Such aggregates are more densely arranged below the embossed channel than adjacent to it, consistent with the significance of compressive forces inferred from our indentation experiments. Renderings of the associated 3D microstructure shown in Fig. S8 and Videos S1-S3 demonstrate that such heterogeneity persists throughout the entirety of each embossed electrode. Therefore, we coarse-grained fine-resolution CT intensity fields using a 3D box filter (MATLAB *imboxfilt3*) of 36 μm size to measure local electrode density by averaging out the heterogeneity from individual electrode constituents. Each coarse-grained CT intensity field possesses a high-density zone (Zone I) directly beneath the embossed channel and a low-density zone adjacent to the channel (Zone II). Further, the degree of densification in Zone I increases as die height increases. The boundary between these zones is also diffuse, and it extends at a 30° to 45° angle from each corner of the channel toward the electrode's current collector. The coarse-grained intensity field for the electrode with greatest die height further shows a degree of current-collector bending that, according to thin-film mechanics theory,[65] is indicative of residual stress in the porous electrode. We posit that this behavior results from the non-uniform, bi-directional plastic flow of electrode material that occurs when it is compressed to its jamming threshold.[66]

Microscopy of the electrode with 200 μm die height confirms the presence of zones with disparate electrode density (Fig. 1f). Regions beneath the embossed channel and along channel surfaces (red boxes) show reduced pore volume with higher binder content and denser particle packing compared to bulk-like electrode regions that occur away from the channels (blue boxes). Collectively, these observations suggest that the compressive stress applied during embossing introduces non-uniform material distribution that sustains porosity, fluid permeability, and ion transport adjacent to channels while constricting them beneath channels. Choosing the extent of channel embossing with which to jointly design die heights and set calendar rolls is critical to producing channels of the desired depth that simultaneously minimize the impacts of densification beneath channels. Using this knowledge, we subsequently design dies to pattern eμ-IDFFs with channels shallower than 200 μm on the present electrodes having 15-17 mg/cm$^2$ areal loading (see SI Sec. 8), while simultaneously calendering them in a single step. Figure 1d shows the height map of PBA electrodes that were embossed with eμ-IDFFs using tapered channels under such conditions.

## *2.2. Tapering of eμ-IDFFs for Uniform Flow with Isometric Flow Paths*

We presently analyze and later implement certain eμ-IDFF generations that are depicted in Fig. 2a. Our past work[5,45] has shown that eμ-IDFFs using channels with a cross-section that is invariant with position $x$ along these channels (i.e., straight channels) produce dead zones of reduced velocity that can arise when the channel-to-electrode hydraulic resistance ratio $\Xi = k_{pe}h_eL^2/sG(x = 0)$ exceeds unity. Here, $k_{pe}$ is electrode hydraulic permeability, $h_e$ is electrode thickness, $s$ is the flow-path length through the porous electrode material between channels, $x$ is position along the eμ-IDFF of interest, and $G^i = -\mu \dot{V}_{\parallel}^i/(dp^i/dx)$ is local hydraulic conductance of channel $i$ ($i = H$ for high-pressure channels and $i = L$ for low-pressure channels). In turn, $\mu$ is dynamic viscosity of the working fluid, $\dot{V}_{\parallel}^i$ is the local longitudinal flow rate within channel $i$, and $dp^i/dx$ is the local longitudinal pressure gradient within that channel. We subsequently improved upon such first-generation straight channels (Gen. I) by using associated quasi one-dimensional (quasi-1D) theory of flow to introduce eμ-IDFFs possessing channels with hydraulic conductance that varies with position, so as to produce uniform transverse fluid velocity. By populating a semi-empirical library that relates hydraulic conductance to discrete channel cross-section geometries, we co-designed a second generation of eμ-IDFFs with piecewise-tapered channels (Gen. II) for electrode patterning by mechanical milling.[45] We presently introduce a third-generation eμ-IDFF (Gen. III) that leverages the enhanced patterning resolution of embossing by incorporating continuous tapering of channel width to produce uniform flow. In conjunction these Gen. III eμ-IDFFs simultaneously are designed to produce isometric (i.e., equal length) flow paths through the porous electrode material that spans between channels.

We now briefly explain the underlying theory by which Gen. III eμ-IDFFs are designed before using quasi three-dimensional (quasi-3D) modeling to show differences in the local flow-field produced among the three eμ-IDFF generations of interest. To enable Gen. III eμ-IDFFs to produce uniform flow first requires that the pressure gradient in each high-pressure channel be identical at a given position $x$ to that in each low-pressure channel:[45]

$$\frac{dp^H}{dx} = \frac{dp^L}{dx}$$

Uniform transverse velocity jointly requires that the longitudinal flow rates in high- and low-pressure channels vary linearly with $x$.[45] By using the definition of channel hydraulic conductance introduced above together with the known distribution of flow rates, the variations of high- and low-pressure channel conductance with position ($G^H(x)$ and $G^L(x)$ respectively) must satisfy the following constraint equation to produce uniform transverse velocity:[45]

$$G^H(x) \cdot x = G^L(x) \cdot (L - x)$$

Eq. 1

While in Ref. [45] we showed that this equation admits an infinite number of conductance functions when high- and low-pressure channels are anti-symmetric – including the linear conductance variation that Gen. II eμ-IDFFs approximate – we instead introduce the additional condition that flow paths be isometric for Gen. III eμ-IDFFs. This condition requires the following constraint on the associated variations of channel width $w^i(x)$ that are admissible for each channel $i$:[67]

$$w^H(x) + w^L(x) = w_0$$

Eq. 2

Equations 1 and 2 are used to determine tapering profiles by recognizing that the laminar, fully developed conductance $G(w,h)$ for Poiseuille flow is principally a function of local channel width $w^i(x)$ and its depth $h$, which is assumed to be constant for Gen. III eμ-IDFFs. $G(w,h)$ is expressed in exact form as an infinite Fourier series following Boussinesq's derivation:[68]

$$G(w,h) = \frac{w^3 h}{12}\left(1 - \frac{192w}{\pi^5 h}\sum_{n=1}^{\infty}\frac{\coth(\pi(2n-1)h/w) - \operatorname{csch}(\pi(2n-1)h/w)}{(2n-1)^5}\right)$$

Eq. 3

However, $G(w,h)$ can also be approximated within 3% of this formula by using second-degree Pade approximant that interpolates between local-cubic[69] and local-linear[70] asymptotes that correspond respectively to the limits of tall, narrow cross-sections ($w \ll h$) and short, wide cross-sections ($w \gg h$):[67]

$$\frac{G(w,h)}{h^3 w/12} \approx \frac{(w/h)^2}{1 + 0.3712(w/h) + (w/h)^2}$$

Eq. 4

Therefore, we can use either Eq. 3 or 4 with the isometric flow-path constraint of Eq. 2 to solve Eq. 1 for the dimensionless position $x^* = x/L$ that corresponds to a certain high-pressure channel width $w^H$:

$$x^* = \frac{x}{L} = \frac{G(w_0 - w^H, h)}{G(w^H, h) + G(w_0 - w^H, h)}$$

Eq. 5

Equation 5 can be used to determine channel profiles by sampling a range of high-pressure channel widths between zero and a maximal width $w_0$, the result of which is shown in Fig. 2a for an eμ-IDFF that we implement later with a depth $h =$ 200 μm and a maximal width $w_0 =$ 514 μm.

For each of the three eμ-IDFF generations, we also performed quasi-3D modeling to capture the effects of incomplete penetration of channels through the thickness of electrodes (see Fig. 1). Here, quasi-3D modeling approximates the pressure and velocity fields in channels as one-dimensional (1D) in the longitudinal ($x$) direction, while approximating fields in the porous electrode as two-dimensional (2D) in the depth-wise ($z$) and width-wise ($y$) directions (see SI Sec. 4). This approach is justified over 3D full-field simulations[71] when electrode length $L$ (45 mm) exceeds the flow-path length $s$ (500 μm) between channels substantially. Our quasi-3D model couples 2D effects in the porous electrode to 1D effects in channels by introducing a transverse conductance $G_\perp$ that relates the unit-length transverse flow-rate $\dot{V}'_\perp$ to the pressure difference $\Delta p = p^H(x) - p^L(x)$ between channels:

$$\dot{V}'_\perp = \frac{G_\perp}{\mu} \Delta p$$

$G_\perp$ results from 2D flow within the porous electrode, the streamlines for which are depicted at representative positions in Fig. 2e for the three eμ-IDFF generations of interest. Because Gen. I and III eμ-IDFFs produce flow paths that do not change substantially along the length of each channel, Fig. 2b shows that their transverse conductance distributions are practically constant. In contrast, Fig. 2b shows that the Gen. II eμ-IDFF produces transverse conductance deviations as large as 12% due to the joint, piecewise variation of its channel width and depth. These variations cause the Gen. II eμ-IDFF to produce deviations of transverse velocity of similar magnitude to its transverse conductance deviation (Figs. 2d,e), while the Gen. III eμ-IDFF produces miniscule deviations from uniformity. On the other hand, Figs. 2d,e show that the Gen. I eμ-IDFF produces deviations from uniformity larger than 100%, because straight channels produce larger pressure differences near the ends of the domain.[5,45] We note additionally that all generations exhibit a region of suppressed velocity beneath channels because of their constricted flow paths along the $y$ direction, likely resulting in incomplete utilization of active material charge capacity in those regions. Future work may therefore target eμ-IDFF co-design for full-depth embossing to eliminate the secondary dead zones that incomplete channel penetration creates.

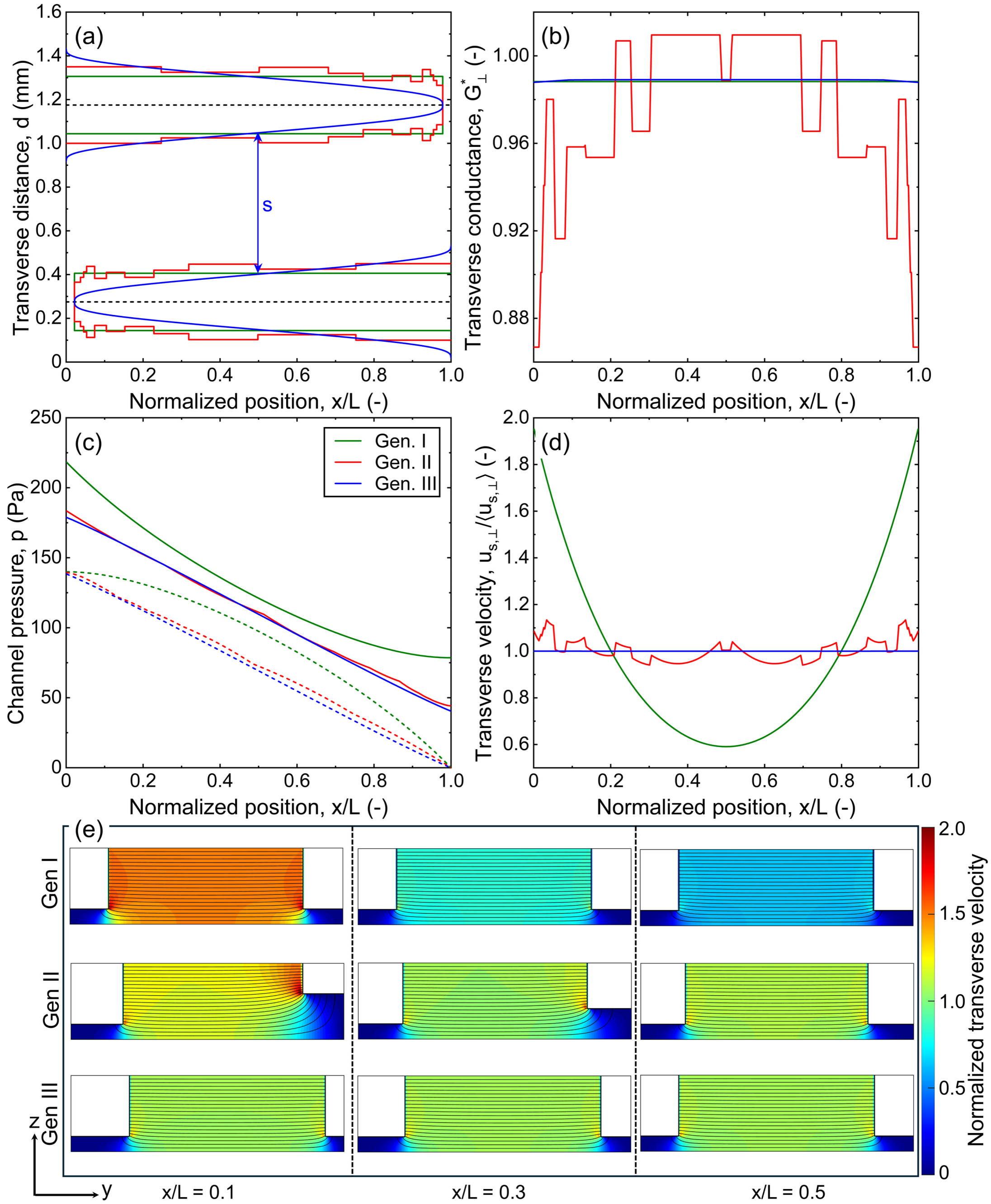

**Figure 2**: Comparison of straight (Gen. I), piecewise, jointly tapered (Gen. II), and isometrically tapered (Gen. III) eμ-IDFF designs. (a) eμ-IDFF channel profiles for the three design generations of interest. (b) Normalized transverse conductance profiles along channel length. (c) Pressure profiles for high-pressure (solid lines) and low-pressure (dashed lines) channels. (d) Normalized transverse velocity distributions. (e) Streamlines within porous electrodes at the normalized positions of interest with contours showing transverse velocity magnitude.

### *2.3. Flow-Cell Characterization of Embossed eμ-IDFFs*

We embossed Gen. II and Gen. III eμ-IDFFs into 45 mm x 45 mm PBA electrodes and performed galvanostatic (i.e., direct current (DC)) electrochemical testing to characterize their specific capacity in a symmetric FDI flow cell using an anion-exchange membrane (AEM) to separate electrodes (Fig. 4a,b). In addition to electrodes embossed with eμ-IDFFs, a pair of unpatterned electrodes served as a control by calendering them to the same thickness as the patterned electrodes. A range of current densities were used between 0.1 and 8 mA/cm$^2$ to jointly assess mechanisms of DC polarization. A flow rate of 1 mL/min was used in each electrode for all tests. Here, the flow cell was plumbed using a single reservoir containing 500 mM aqueous NaCl that both electrodes simultaneously drew from and discharged into. Under such a condition salt concentration in the associated reservoir remains practically unchanged during cycling, enabling the isolation of electrode-induced ohmic and concentration polarization from polarization caused by desalination.

Regardless of the eμ-IDFF architecture used, all electrodes showed a monotonic decrease in specific capacity with increasing current density (Fig. 3a) because of the finite potential window used for cycling. This capacity drop was substantially greater for unpatterned electrodes due to their large polarization (Fig. 3b). In turn, Gen. III-embossed electrodes produced highest capacity at large current densities. This improvement is attributable in part to the uniform flow distribution produced by the isometric flow paths of Gen. III electrodes (see Fig. 2d,e). However, Gen. II electrodes showed specific capacity at parity with Gen. III electrodes at low current because the driving force for streamwise state-of-charge gradients was reduced by the decreased salt concentration swing that results from decreasing current while maintaining flow-rate parity.[72–74] When cycling at 1 mA/cm$^2$ after cycling at higher rates, all patterned electrodes nearly recovered their initial capacity, indicating that the capacity drop observed at high rates was not due to electrode degradation. Although unpatterned electrodes also showed some capacity recovery, their capacity remained below those of embossed electrodes.

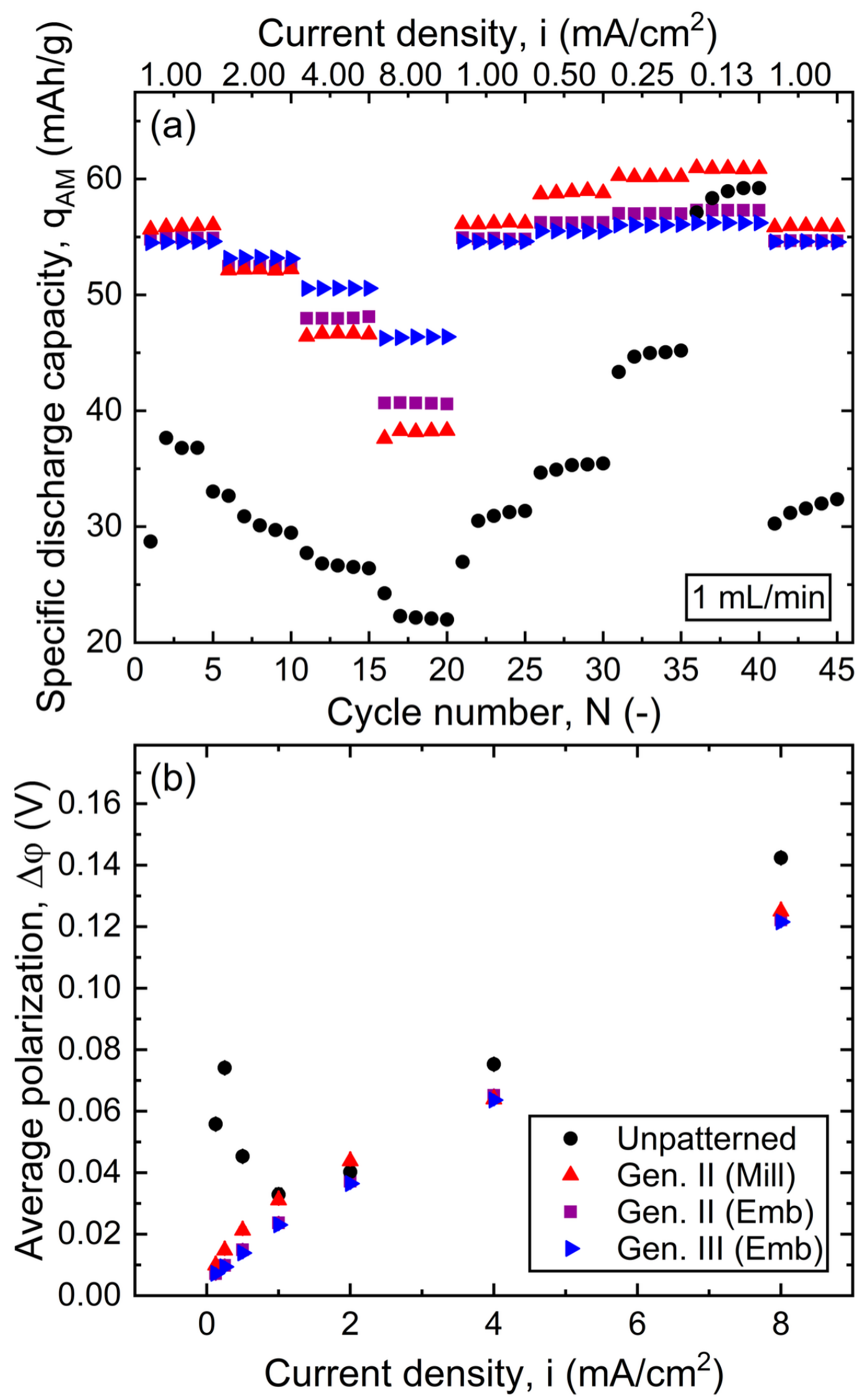

**Figure 3:** (a) Capacity retention and (b) average polarization during galvanostatic cycling at various rates for electrodes embossed with Gen. III eµ-IDFFs and either embossed or milled with Gen. II eµ-IDFFs. Unpatterned electrodes of the same thickness served as a control. All electrodes used 1 mL/min flow rate of 500 mM NaCl solution drawn from and discharged into a common reservoir.

These capacity variations are accompanied by certain variations of average polarization with current density (Fig. 3b). All patterned electrodes showed a linear increase in average polarization with current density, consistent with ohmic response, whereas unpatterned electrodes displayed a non-linear dependence in the low-rate regime. This non-linear behavior suggests that the irregular, multi-scale pore topologies that unpatterned electrodes possess[77] may produce non-uniform flow distribution. Embossing of electrodes with eμ-IDFFs thus regularizes a given electrode's macroscopic porosity and, consequently, the flow produced through it. These trends were consistently observed at 5 mL/min (Fig. S10), further establishing the robust cycling of electrodes patterned with tapered-channel eμ-IDFFs across flow conditions. We next evaluate the performance of PBA electrodes with these eμ-IDFF generations in a symmetric FDI flow cell for desalination with seawater- and brackish-salinity feeds.

### *2.4. Desalination by FDI using Embossed eμ-IDFFs*

Desalination experiments using PBA electrodes patterned with Gen. II and III eμ-IDFFs were performed using the symmetric FDI flow cell already described (Fig. 4a,b). In contrast with the aforementioned electrochemical characterization under flow using a single reservoir, the present experiments were conducted in batch mode using separate reservoirs in which diluate and brine were generated while being recirculated through the flow cell to minimize streamwise concentration gradients.[75,76] A flow rate of 1 mL/min was used in all experiments to ensure direct comparison between electrodes with different eμ-IDFFs being cycled at different current densities, while limiting cell potential between -0.4 V and 0.4 V.

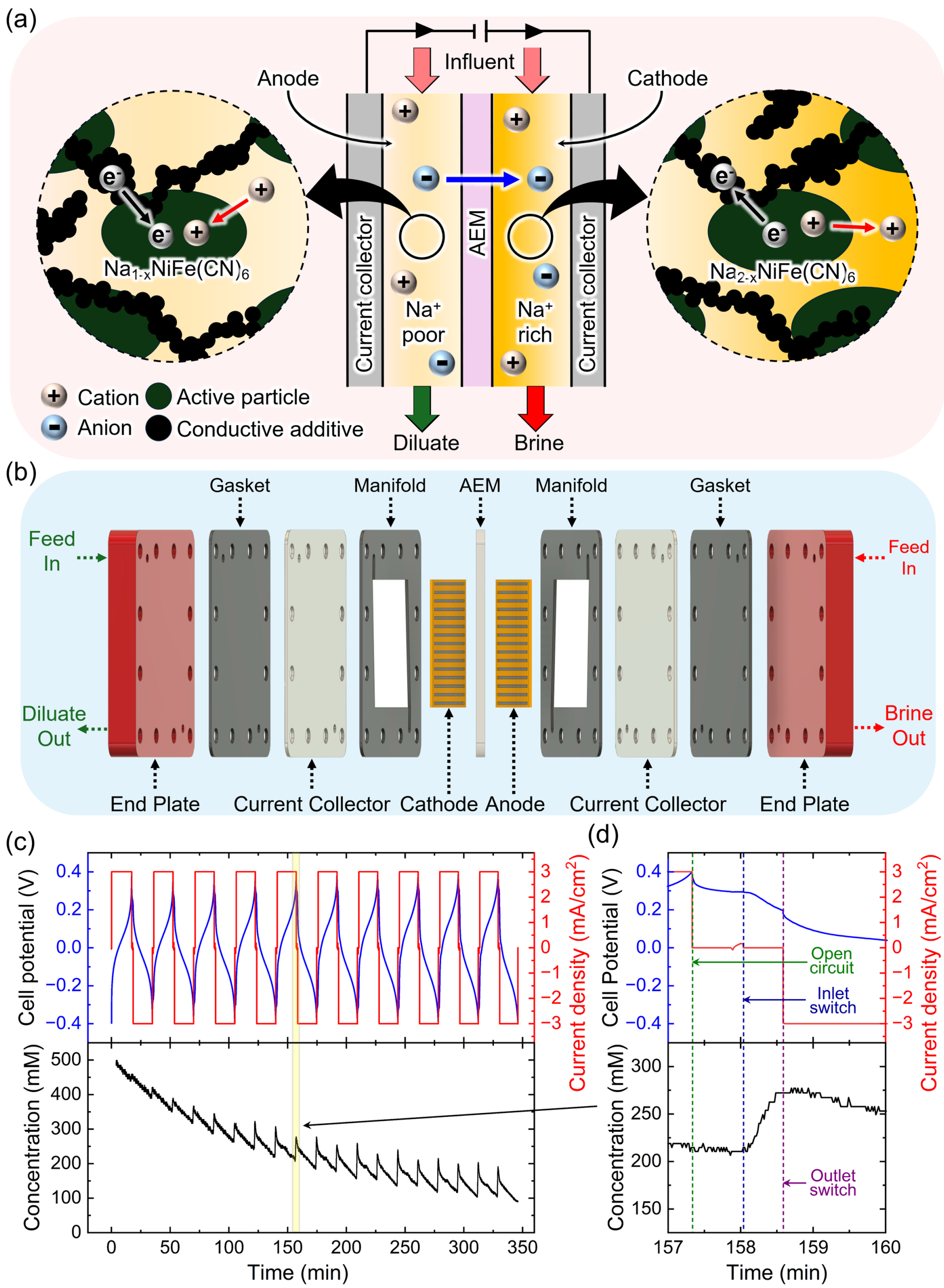

**Figure 4:** (a) Architecture of the symmetric Faradaic deionization flow cell used. (b) Exploded view of the flow cell incorporating PBA electrodes patterned with eμ-IDFFs. (c) Time-resolved profiles of applied current, cell potential, diluate concentration during successive galvanostatic charge and discharge half-cycles with 500 mM NaCl influent for electrodes embossed with Gen. III eμ-IDFFs. (d) Open-circuit interval followed by inlet-outlet valve switching sequence after each half-cycle. An optimized delay time is introduced between valve-switching events to minimize apparent mixing volume.

The temporal variations of applied current, cell potential, and diluate concentration during successive charge/discharge half-cycles are shown in Fig. 4c using electrodes embossed with Gen. III eμ-IDFFs. During each half-cycle, one electrode acts as the cathode, intercalating cations during PBA reduction, while anions migrate through an AEM to maintain charge neutrality, producing a diluate stream. Simultaneously, the counter electrode acts as the anode, deintercalating cations into water and producing a brine stream during PBA oxidation. At the end of each half-cycle, the former electrode approaches 100% state-of-charge (SOC) while the later approaches 0% SOC. The applied current is then switched off and the FDI cell enters open-circuit (OC) mode, during which inlet and outlet valves are switched to reverse flow direction, followed by inversion of current polarity to ensure continuous desalination (Fig. 4d).[77] To minimize the apparent intermixing between brine and diluate streams, an optimized delay time of 33 s was introduced between valve-switching events (Fig. 4d), determined from breakthrough curve analysis (see SI Sec. 10). The associated breakthrough curves for Gen. III eμ-IDFFs possess a longer median residence time than Gen. II eμ-IDFFs, consistent with our predictions in Fig. 2 that Gen. III possesses fewer dead zones than Gen. II. Despite the benefits of eμ-IDFF optimization, apparent intermixing becomes increasingly significant at higher concentration differences between the brine and diluate reservoirs when the concentration change induced by intermixing approaches the salt concentration change removed electrochemically within a half-cycle.[5]

Figure 5 shows the resulting desalination performance of electrodes patterned with Gen. II and Gen. III eμ-IDFFs using a feed with seawater level salinity as 500 mM NaCl. Generation II eμ-IDFFs were patterned both by milling and embossing, whereas the Gen. III eμ-IDFFs were patterned by embossing alone. All experiments were conducted in triplicate, with reported data representing the mean and with error bars indicating their standard deviation. Despite differences in eμ-IDFF geometry, a comparable level of salt removal (≈96%) was achieved across all designs, reaching near-freshwater concentration (≈17 mM) with final concentrations of approximately 23 mM for Gen. II-milled, 20 mM for Gen. II-embossed, and 21 mM for Gen. III-embossed electrodes (Fig. 5a). Notably, Gen. III-embossed electrodes reached the same level of salt removal in less desalination time compared to both Gen. II electrode types, reflecting differences in charge efficiency.

Generation II electrodes showed similar charge efficiency whether patterned by milling or embossing, and Gen. III electrodes demonstrated 23% improvement over them both (Fig. 5b). This enhancement in charge efficiency is attributed to more uniform flow distribution enabled by isometric flow paths of Gen. III eμ-IDFFs, which improves electrolyte accessibility and maximizes active material utilization while minimizing concentration gradients. These gains in charge efficiency directly translate into reduced energy consumption. Although Gen. III electrodes incur approximately 50% higher pumping energy compared to Gen. II electrodes, their volumetric energy consumption (including pumping and electrical energy) is reduced by approximately 16% (Fig. 5c). Furthermore, Gen. III-embossed electrodes achieved 1.5-fold improved total thermodynamic energy efficiency (TEE) and 1.5-fold improved water recovery at $\approx$ 96% salt removal as compared to Gen. II electrodes (Fig. 5d and SI Sec. 11). At 40% salt removal, total TEE reached approximately 36%, which is roughly double that which we reported previously at comparable salt removal,[45] owing to reduced pumping losses at the low flow rate of 1 mL/min used here.

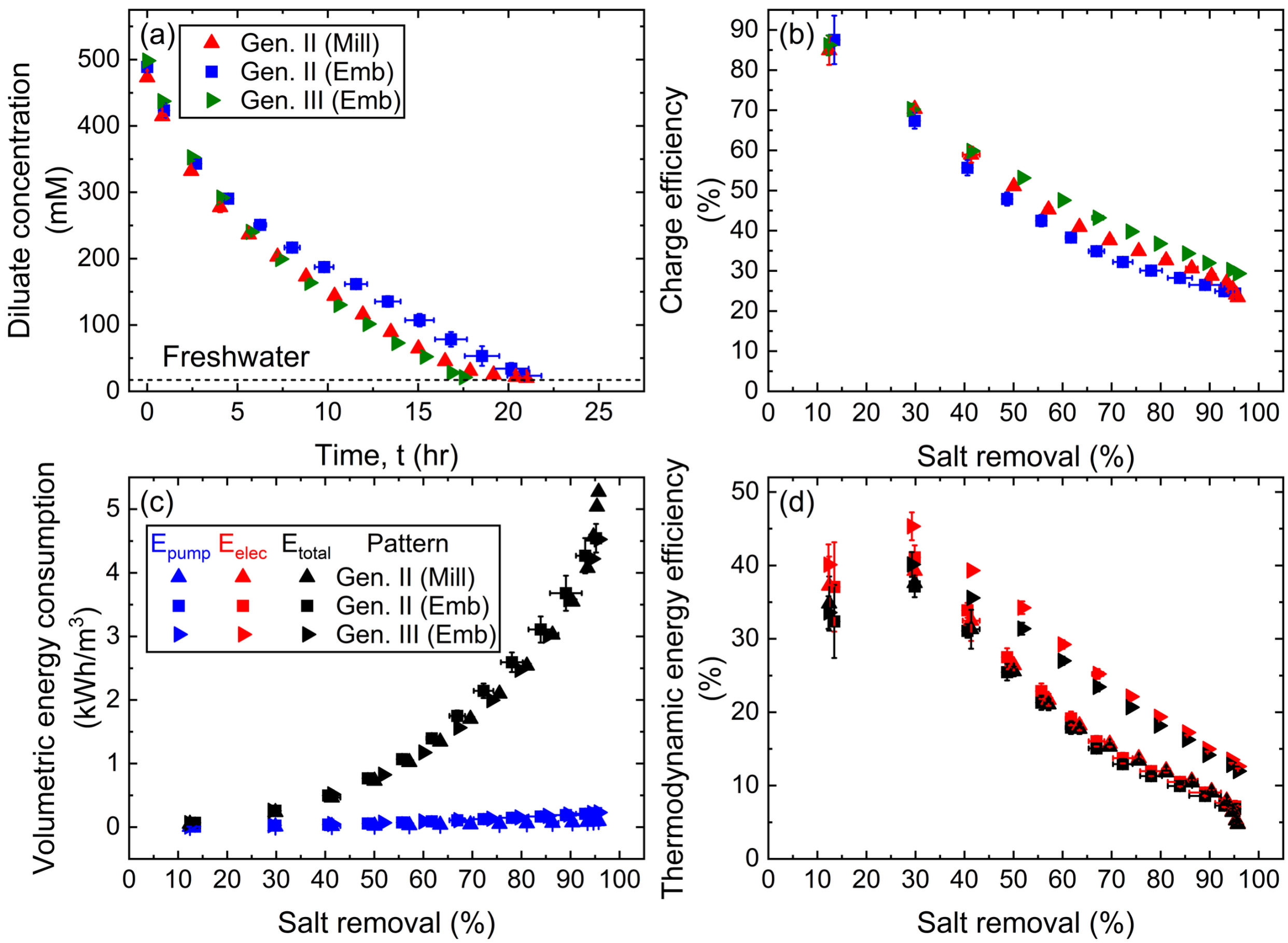

**Figure 5:** Desalination performance of PBA electrodes patterned with Gen. II (milled and embossed) and Gen. III (embossed) eμ-IDFFs using a NaCl seawater-salinity feed (SW, 29.2 g/L ≈ 500 mM NaCl): (a) diluate salt concentration, (b) charge efficiency, (c) volumetric energy consumption, and (d) thermodynamic energy efficiency.

Figure 6 shows desalination performance of the same Gen. II and III electrodes, but now using a 100 mM NaCl brackish water feed. Here, Gen. III eμ-IDFFs are shown to achieve a diluate concentration of ≈10 mM at 90% salt removal with a total energy consumption of only 0.32 $kWh/m^3$ including pumping and electrical energy (Fig. 6a,c). This corresponds to a 30% reduction in energy compared to Gen. II electrodes. As with seawater-salinity results, this improvement is attributed to the isometric flow paths that are particular to Gen. III eμ-IDFFs that promote uniform transverse velocity between high- and low-pressure channels. Enhanced flow uniformity results in more uniform current distribution, yielding a 15% increase in charge efficiency (Fig. 6b). Consequently, thermodynamic energy efficiency increases to approximately 1.5 times that of Gen. II electrodes at 90% salt removal (Fig. 6d).

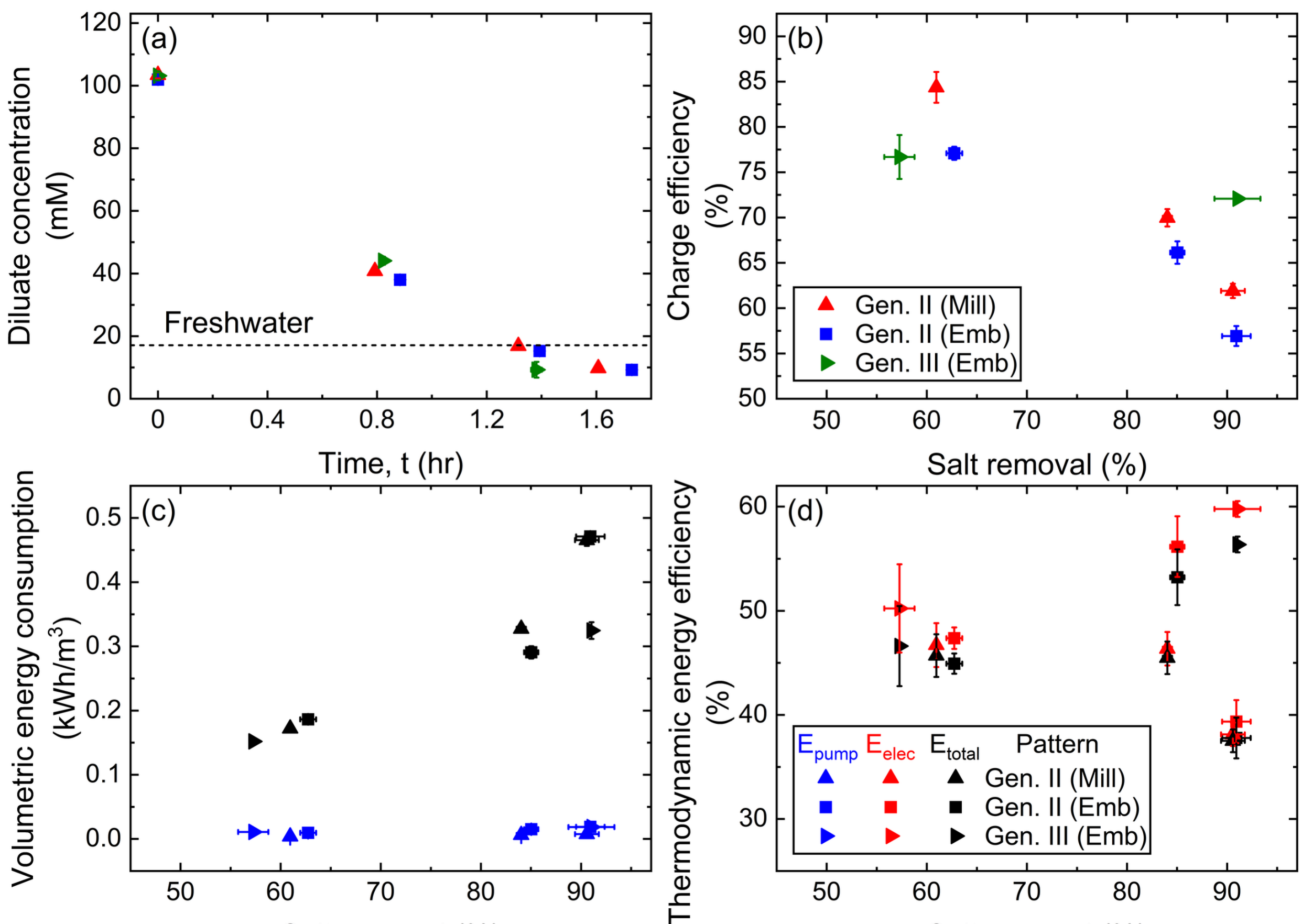

**Figure 6:** Desalination performance of PBA electrodes patterned with Gen. II (milled and embossed) and Gen. III (embossed) eμ-IDFFs using a NaCl brackish-salinity feed (BW, 5.84 g/L ≈ 100 mM NaCl): (a) diluate salt concentration, (b) charge efficiency, (c) volumetric energy consumption, and (d) thermodynamic energy efficiency.

The salt removal and energy consumption achieved when desalinating with embossed Gen. III eμ-IDFFs stand out among the literature. Molar specific energy consumption (SEC) was determined for Gen. III-embossed electrodes and benchmarked against prior FDI studies using PBA electrodes in symmetric FDI (Fig. 7). Previous studies have generally reported either modest salt removal for brackish feeds [78–81] or relatively low salt removal for seawater-salinity feeds. [82] Our earlier work[5,77] improved salt removal for both feeds using laser-machined Gen. I eμ-IDFFs, though at comparatively high energy consumption due to material loss during patterning and low permeability. Subsequently, Gen. II-milled electrodes showed salt removal exceeding 90%, with SEC of 26 kJ/mol for brackish feeds and 54 kJ/mol for seawater feeds.[45] In contrast, the present Gen. III-embossed electrodes show SEC of 13 kJ/mol for brackish feeds and 34 kJ/mol for seawater feeds, corresponding to reductions of about 50% and 37%, respectively. These results demonstrate that the isometric channel design offered by Gen. III eμ-IDFFs substantially improves desalination performance while lowering energy consumption across both feed conditions. Furthermore, the SEC required to achieve varying levels of salt removal is consistently lower than other past studies report.[77–82]

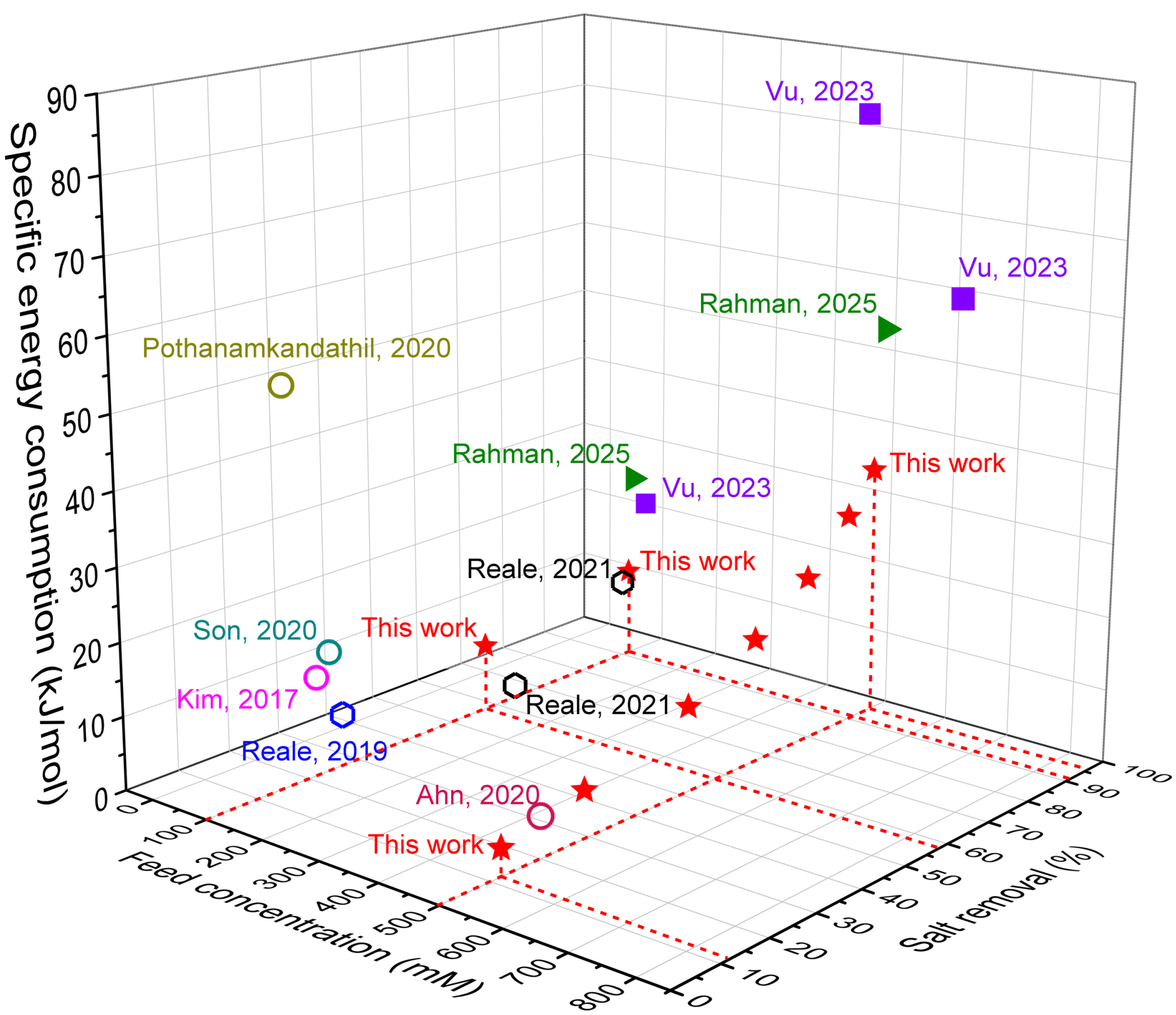

**Figure 7:** Comparison of specific energy consumption, as a function of feed concentration and salt removal, with previous studies[5,45,77–82] using symmetric FDI. Unfilled symbols include electrical energy alone and filled symbols include pumping and electrical energy.

Generation III eμ-IDFFs are additionally shown to increase water production rate. Desalination with electrodes embossed with Gen. III eμ-IDFFs, and with Gen. I and II eμ-IDFFs in our past work,[5,45] shows relatively low freshwater production rates of 0.14 L/m$^2$-h for seawater and 2.81 L/m$^2$-h for brackish water desalination at 1 mA/cm$^2$, as compared to electrodialysis (≈10 L/m$^2$-h).[83,84] To improve productivity, desalination experiments were conducted at current densities of 2 and 3 mA/cm$^2$ with seawater-salinity feed (Fig. 8). In all cases, the feed concentration was ≈500 mM NaCl, and the targeted diluate concentration was ≈100 mM NaCl, corresponding to brackish salinity. As shown in Fig. 8a, water production rate increases with increasing current density, albeit in a non-linear fashion. Increasing current density from 1 to 2 mA/cm$^2$ doubled the production rate from 0.24 to 0.49 L/m$^2$-h, while maintaining approximately 80% salt removal. This productivity gain is attributed primarily to faster salt removal and shorter cycle duration rather than to a greater extent of salt removal, as reflected in the diluate concentration profiles shown in Fig. 8b. Further increasing to 3 mA/cm$^2$ yielded only ≈17% gain in production rate (0.58 L/m$^2$-h). This smaller productivity gain suggests that the benefit of increasing current density becomes less effective at higher operating currents. Electrical energy consumption also increased monotonically, while pumping energy remained nearly constant (Fig. 8c). The higher electrical energy demand at 3 mA/cm$^2$ is consistent with increased polarization (Fig. 3) and increased parasitic charge storage, such as electrostatic co-ion adsorption, at elevated current densities. Correspondingly, the charge efficiency decreased with increasing current density, indicating that a smaller fraction of the electrons transferred contributed directly to salt removal (Fig. 8d). TEE also declined, reflecting the greater electrical energy input required to achieve comparable desalination performance (Fig. 8e). Although water recovery remained consistently higher over repeated cycles at higher current densities – a consequence of shorter cycle durations from faster ion removal (Fig. 8f) – this gain was accompanied by higher electrical energy consumption and lower charge and thermodynamic energy efficiency, highlighting a trade-off between productivity and energy.  Future work may focus on optimizing rate capability, mitigating apparent intermixing, and scaling up the embossing process to fully realize the potential of isometric, tapered eμ-IDFFs in practical, large-scale electrochemical desalination systems.

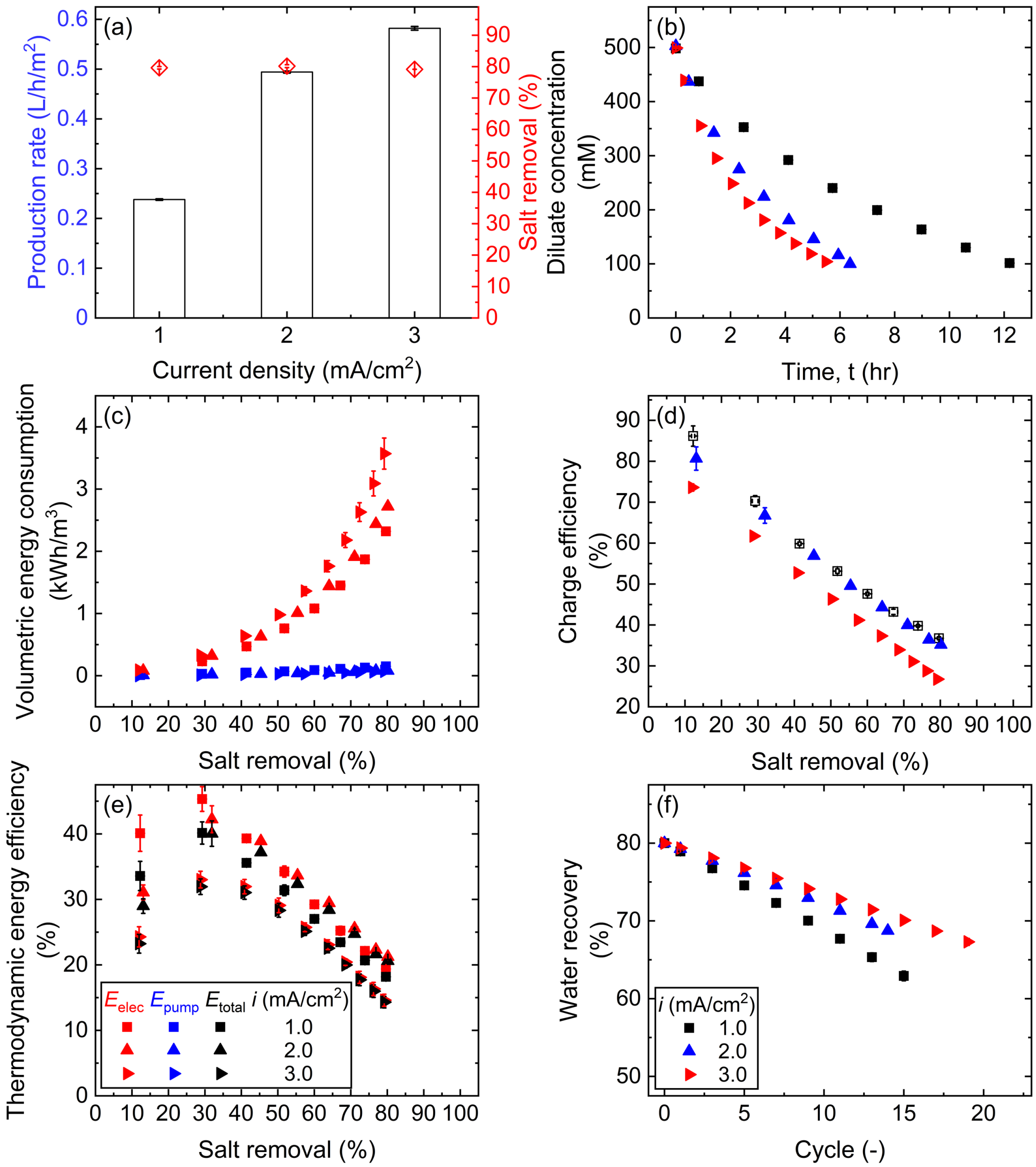

**Figure 8:** Effects of current density on desalination performance using PBA electrodes embossed with Gen. III eμ-IDFFs with feeds having seawater salinity as 500 mM NaCl. (a) Production rate and salt removal as a function of applied current density. (b) Time variation of diluate concentration. (c) Total volumetric energy consumption, (d) charge efficiency, and (e) thermodynamic energy efficiency as a function of salt removal. (f) Water recovery as a function of cycle number.

## 3. Conclusions

We introduce a non-subtractive embossing method that is also scalable and high throughput to pattern flow fields onto porous electrodes. This method precisely resolves channel features while preserving active material and reducing fabrication time by approximately 100-fold relative to mechanical milling. Indentation experiments indicate that embossing permanently forms channels in porous electrodes with limited elastic recovery, enabling their precise implementation. The observed linear relationship between normal pressure and the extent of permanent material displacement was used to create a mechanical model which shows that embossing suppresses the required force to pattern channels when electrode length is up-scaled. Embossing produces such force because contact length $x_c = \sqrt{2R\Delta h}$ is determined by roller radius $R$ and thickness change $\Delta h$, rather than by the electrode's length $L$ as in stamping. In turn, plastic compression is shown to preserve open porosity within the microstructure adjacent to channels during plastic compression, though it densifies electrode material beneath channels as the die height increases. Die height should therefore optimize channel depth and local densification by tuning feature size and roller gaps to electrode thickness prior to embossing and calendering.

We also introduce a novel embedded, micro-interdigitated flow field (eμ-IDFF) design that simultaneously tapers channels in their width while maintaining equal spacing between adjacent high- and low-pressure channels. Ultimately, we emboss Prussian blue analogue electrodes with this eμ-IDFF in order to perform electrochemical desalination. We use quasi one-dimensional theory of pressure-driven flow through the eμ-IDFF to define the specific width tapering profile that jointly satisfies flow uniformity and isometric flow path constraints. The particular width profile used here for a certain maximal-cross-section's aspect ratio belongs to a family of width profiles, as determined by using Eq. 5 together with an appropriate formula for the dependence of hydraulic conductance of channel width and depth. The channel conductance formulas that we have used thus far (Eq. 3 or 4) are for laminar, fully developed flow, and operation at higher flow rates could produce inertial effects in the resultant flows. Variations of electrode permeability from the constant value assumed in our analysis could also arise from either a flow-path dependence of permeability[45] or local densification. Finally, eμ-IDFF design criteria can be expanded to include a desired residence time distribution to minimize interstream mixing.

We also demonstrate that coupling flow-field design with scalable electrode patterning can substantially improve desalination performance of porous intercalation electrodes in Faradaic deionization. Desalination results obtained using electrodes embossed with eμ-IDFFs show competitive performance relative to past demonstrations of Faradaic deionization using Prussian blue analog electrodes. Tapered, isometric eμ-IDFFs also showed higher capacity retention and lower polarization than unpatterned electrodes, particularly at high current densities, consistent with their more uniform flow distribution. These gains project onto desalination performance across both seawater and brackish water salinity feeds. With 500 mM NaCl feeds, tapered, isometric eμ-IDFFs achieved 96% salt removal in less time than eμ-IDFFs using piecewise, joint variation of width and depth, with higher charge efficiency, lower volumetric energy consumption, and higher thermodynamic energy efficiency. With 100 mM NaCl feeds, the tapered, isometric eμ-IDFF reached 90% salt removal using only 0.32 kWh/m$^3$ including both electrical and pumping energy. The resulting specific energy consumption of 34 and 13 kJ/mol for seawater- and brackish-salinity feeds, respectively, is lower than that reported in prior symmetric FDI studies. With seawater-salinity feeds, doubling current density from 1 to 2 mA/cm$^2$ also doubled production rate, but a further increase to 3 mA/cm$^2$ yielded only a 17% gain at the cost of higher electrical energy and lower charge efficiency. This diminishing behavior is attributed to the faster charge/discharge cycling together with local salt depletion inside the porous electrodes at low flow rates. Future work should focus on high flow rates to mitigate local salt depletion while balancing the added pumping energy, and determining the current density and potential window at which side reactions begin.

Overall, this work establishes an integrated embossing-calendering method as a scalable route to manufacturing flow-engineered porous electrodes, in which flow-field design and patterning methods should be considered together. Future work may address full-depth embossing to eliminate the secondary dead zones that incomplete channel penetration creates beneath embossed channels, engraved calender rolls for continuous roll-to-roll fabrication, and larger electrode areas and stacked electrode pairs to improve water production rates while retaining the energy efficiency benefits demonstrated here. Testing with realistic feeds that include natural seawater and brackish groundwater containing a diversity of salts, dissolved organics, and suspended particulates should be tested to evaluate ion selectivity, scaling behavior, and the long-term stability of embossed electrodes. Beyond desalination embossed eμ-IDFFs could benefit other technologies where electrolyte distribution limits performance, including redox-flow batteries, flow-through electrochemical reactors, and selective ion recovery systems such as lithium extraction.

## 4. Materials and Methods

### *4.1. Synthesis of PBA Nanoparticles*

Nickel hexacyanoferrate PBA nanoparticles were synthesized using a co-precipitation technique as described in our previous studies.[5,45,85] Equal volumes of 100 mM potassium ferricyanide [$K_3Fe(CN)_6$] and 200 mM nickel (II) chloride hexahydrate [$NiCl_2.6H_2O$] solutions were added dropwise into a beaker containing deionized (DI) water under vigorous stirring at room temperature. A peristaltic pump (Masterflex 7523-90, Cole Parmer) was used to maintain a controlled flow rate of 1.5 mL/min for both solutions, ensuring uniform mixing, and resulting in a brownish-yellow dispersion of about 1 wt% PBA nanoparticles. The dispersion was subsequently sonicated for 30 min and aged overnight at room temperature to achieve uniform particle size distribution. The nanoparticles were then collected by centrifugation (Eppendorf 5800R, Fisher Scientific) at 3900 rpm for 5 min. This process was repeated several times until the supernatant solution becomes clear, ensuring the removal of unreacted precursors and byproducts. The final product was dried overnight in an electric oven at 70°C.

### *4.2. Fabrication of Porous Electrodes*

The porous electrodes used in this study were fabricated from a composite consisting of PBA nanoparticles as active material, Ketjen Black (EC-600JB ) as a conductive additive, and polyvinylidene fluoride (PVdF, Solvay Solef 5130) as a binder in an 80:5:15 mass ratio. The PVdF binder was first dissolved in N-methyl-2-pyrrolidone (NMP, Sigma-Aldrich) at a concentration of 50 mg/mL using a planetary mixer (Thinky ARE-310) operated at 2000 rpm for 25 min followed by a 3-min degassing at 2200 rpm to obtain a viscous solution. Meanwhile the PBA nanoparticles and conductive additive were ground together with a mortar and pestle, transferred to a grinding tube containing 18-20 stainless steel balls (5 mm diameter), and further processed in a vortex mill (Ultra Turra-X, IKA) at 6000 rpm for 30 min to obtain a fine, homogenous powder. This powder was then combined with the binder solution at a solvent-to-solid ratio of 3 mL NMP per 1 g to total solids and subsequently homogenized in the planetary mixer at 2000 rpm for an additional 30 min. The resulting slurry was cast onto 120 μm thick graphite foil (Ceramaterials) current collectors using a doctor blade (MTI-XTL) and a film applicator (Elcometer 4340, Elcometer Inc) to a wet film thickness of approximately 1.1 mm. The cast electrode films were solidified by wet-phase inversion, in which they were immersed active-layer-down in a neutral-pH coagulation bath held at 85°C for 10-15 min.[45] This process prevents cracking during solidification,[85] eliminates PVdF dehydrofluorination that occurs at high pH and ensures enhanced electronic conductivity through uniform thermal exposure. The films were then dried in a fume hood for 2-3 h, followed by drying in the electric oven at 70°C for 12 h. The dry electrode thickness, excluding the current collector, was measured to be 265 μm. These electrodes were calendered to 200 μm using a roll-press (MTI-XTL) to improve electrical conductivity and rate capability,[77] and were than patterned with Gen. II eμ-IDFFs over a 45 mm x 45 mm area by mechanical milling. Details of the milling process and corresponding channel characterization are given in our previous work.[45] In the present study, we introduce an integrated embossing-calendering method in which uncalendared electrodes are patterned with Gen. II and Gen. III eμ-IDFFs while simultaneously calendering them as described below.

### *4.3. Embossing of eμ-IDFFs into PBA Electrodes*

Prussian blue analog electrodes were embossed with both Gen. II and Gen. III eμ-IDFFs using dies milled with the negative shape of the eμ-IDFFs (Fig. 1b). The die toolpaths were designed in Autodesk Fusion 360, and the corresponding G-code was executed using a desktop computer numerical control (CNC) milling machine equipped with precision ball screws and closed-loop stepper motors, operated via GRBL Mega v1.1 motion control firmware.[86] A 500 μm diameter square end mill (MSC Industrial Direct) was used to machine the negative shape of the eμ-IDFFs into a 2 mm thick acrylic sheet (McMaster-Carr), mounted on the CNC bed using a perforated aluminum plate (McMaster-Carr). Milling was performed at a spindle speed of 18,000 rpm, and a HEPA-filtered dust collector was used to extract debris during machining. Each die was then passed through the roll-press (MTI-XTL) together with the electrode, backed by a second 2 mm thick acrylic sheet, to emboss the eμ-IDFFs into the electrode while simultaneously calendering it. The roller gap was set for each die according to the channel depth and was defined as the combined thickness of the die, the target electrode film, the current collector, and the backing sheet. A 5° draft angle was added to the channel sidewalls to facilitate release of the electrode after embossing-calendering, minimizing frictional shear on the channel walls and preventing tearing or distortion of the embossed features. The resulting channel cross-sections, measured by optical profilometry (Keyence VK-X1000), were within ±10% of the nominal dimensions (see SI Sec. 2). The internal microstructure of the embossed electrodes was characterized by X-ray CT and SEM. CT scans were acquired on a Zeiss Xradia Versa 630 (Fig. S8) at 20x magnification over a field view of about 0.548 $mm^2$. Data were collected with a binning factor of 2, in which 2 x 2 block of pixels was averaged into a single pixel intensity value to reduce scan time and dataset size. Table S3 summarizes the scan settings for the samples with different degrees of embossing shown in Fig. 1e.

### *4.4. Symmetric FDI Flow Cell and Desalination Experiments*

All desalination experiments were performed using a custom-built symmetric FDI flow cell (Fig. 4a,b). The cell consists of two 45 mm x 45 mm PBA electrodes, acting as cathode and anode, separated by a 150 μm thick AEM (Neosepta AMX), with two 250 μm thick graphite foil sheets (Ceramaterials) as current collectors. EPDM rubber sheets (1 mm thick, McMaster-Carr) were placed between the end plates and the current collectors as seals. Two thinner (400 μm) EPDM gaskets, each integrated with a Z-type manifold, supplied brine and diluate to the respective electrodes. The Z-type manifold design maintains a constant streamwise pressure gradient across the transverse direction, allowing uniform flow distribution through the porous electrodes. Prior to flow-cell assembly, the electrodes were charged and discharged in flooded three-electrode cells containing 500 mM NaCl, with an Ag/AgCl reference electrode (Pine Research, 0.199 V vs. NHE) and a graphite rod counter electrode (StonyLab). Cycling was done at approximately C/10 (2 mA) until one electrode reached 0% state-of-charge (SOC; 0.6 V vs. Ag/AgCl) and the other reached 100% SOC (0.1 V vs. Ag/AgCl). Desalination experiments were conducted at 1C (20.25 mA) using a VMP-3 multichannel potentiostat (BioLogic). For feeds with seawater and brackish water level salinity, 2 mL of brine and 8 mL of diluate were recirculated using two peristaltic pumps (NE-9000, New Era Pump System Inc.) at a constant flow rate of 1 mL/min. Salt concentrations were measured using a Compact Ion Chromatography system (Metrohm) and real-time conductivity of the brine and diluate streams was recorded using custom flow-through conductivity probes based on commercial micro-USB connectors and read with a Conduino system.[87]

The recorded conductivity data were then post-processed to obtain real-time salt concentrations in both streams.

### *4.5. Desalination Performance Metrics*

The charge efficiency[88] (CE) was calculated as the ratio of the moles of salt removes from the diluate ($n_{salt}$) to the moles of electrons transferred ($Q/F$), where $Q$ is the charge and $F$ is the Faraday constant. The volumetric energy consumption (VEC, in kWh/m$^3$) was calculated as the total energy (electrical and pumping) required per mole of salt removed in each half-cycle, multiplied by the cumulative salt removal ($SR = \Delta C = C_F - C_D$) at the end of that half-cycle, where $C_F$ and $C_D$ are the feed and diluate concentrations, respectively.

$$VEC = \frac{\int_{t_0}^{t_{end}} I(t)V(t)dt + P_{pump}(t_{end} - t_0)}{n_{salt}} \times SR$$

where $I(t)$ and $V(t)$ are the applied current and cell potential, respectively, as functions of time. The pumping power $P_{pump}$ is the hydraulic power required to maintain a constant volumetric flow rate $(\dot{V})$ through the electrodes and is calculated as $P_{pump} = \dot{V}^2 R_h$; where $R_h$ is the total hydraulic resistance of the electrode, obtained from Darcy's law as $R_h = \mu L / k_h A_e = \Delta p / \dot{V}$. Here, $\Delta p$ is the pressure difference across the porous electrode, $\mu$ is the dynamic viscosity of the working fluid, $k_h$ is the hydraulic permeability of the electrode, $L$ is the electrode length, and $A_e = w_e t_e$ is the cross-sectional area of the electrode normal to flow, where $w_e$ and $t_e$ are the electrode width and thickness.

The thermodynamic energy efficiency (TEE) was calculated as the ratio of the minimum volumetric energy consumption ($VEC_{min}$) to the cumulative VEC at each half-cycle. $VEC_{min}$ is defined as the reversible work required per mole of salt removed in each half-cycle, multiplied by the cumulative salt removal at the end of that half-cycle.[45]

## CRediT Authorship Contribution Statement

**Md Habibur Rahman:** Writing – review & editing, Writing – original draft, Methodology, Investigation, Formal analysis, Data Curation. **Sadia Afrin:** Formal analysis, Writing – review & editing. **Kyle C. Smith:** Supervision, Resources, Project administration, Funding acquisition, Writing – review & editing, Writing – original draft, Methodology, Investigation, Formal analysis.

## Acknowledgements

The authors acknowledge support from the Expeditionary Energy Program of the U.S. Office of Naval Research (Award No. N-00014-25-1-2554) and the Department of Mechanical Science and Engineering at the University of Illinois Urbana-Champaign. Optical Profilometry and Scanning Electron Microscopy were conducted at the Materials Research Laboratory, and X-ray computed tomography were performed at the Beckman Institute for Advanced Science and Technology at UIUC. The authors also thank Neil Pearse for facilitating access to the Trotec Speedy Flexx 400 laser at the Siebel Center for Design at UIUC.

## Declaration of Competing Interest

Kyle C. Smith reports that financial support provided by the U.S. Office of Naval Research. Kyle C. Smith, I. C. Loud IV, Md Habibur Rahman, and Md Abdul Hamid have a pending International Patent Application (PCT/US26/10636). The remaining authors declare no known competing financial interests or personal relationships that could have influenced the work reported in this paper.

## Data Availability

Data will be made available on reasonable request.

## Appendix: Supplementary Materials

Supplementary material associated with this article can be found in a separate document.

## References

(1) Rocha, C.; Knöri, T.; Ribeirinha, P.; Gazdzicki, P. A Review on Flow Field Design for Proton Exchange Membrane Fuel Cells: Challenges to Increase the Active Area for MW Applications. *Renewable and Sustainable Energy Reviews* **2024**, *192*, 114198. https://doi.org/10.1016/j.rser.2023.114198

(2) Liu, B.; Jacquemond, R. R.; Muñoz-Perales, V.; Buzzi, S.; Hjelm, J.; Forner-Cuenca, A. Integrating Flow Field Geometries within Porous Electrode Architectures for Enhanced Flow Battery Performance. *Small* **2026**, *22* (1), e11327. https://doi.org/10.1002/smll.202511327

(3) Lu, M. Y.; Yin, C.; Ma, Q.; Su, H. N.; Lu, P.; Dai, Z. Q.; Yang, W. W.; Xu, Q. Flow Field Structure Design for Redox Flow Battery: Developments and Prospects. *J. Energy Storage* **2024**, *95*, 112303. https://doi.org/10.1016/j.est.2024.112303

(4) Liu, R.; Tian, Y.; Zhang, Q.; Shen, Y. Enhanced Desalination Performance of Battery Electrode Deionization Based on the Optimized Design of Flow Channel Structure and Channel Section Reconstruction. *Ind. Eng. Chem. Res.* **2024**, *63* (30), 13078–13089. https://doi.org/10.1021/acs.iecr.4C01284

(5) Do, V. Q.; Reale, E. R.; Loud, I. C.; Rozzi, P. G.; Tan, H.; Willis, D. A.; Smith, K. C. Embedded, Micro-Interdigitated Flow Fields in High Areal-Loading Intercalation Electrodes Towards Seawater Desalination and Beyond. *Energy Environ. Sci.* **2023**, *16* (7), 3025–3039. https://doi.org/10.1039/d3ee01302b

(6) Pan, Z.; Yang, J.; Kong, J.; Loh, X. J.; Wang, J.; Liu, Z. "Porous and Yet Dense" Electrodes for High-Volumetric-Performance Electrochemical Capacitors: Principles, Advances, and Challenges. *Advanced Science* **2022**, *9* (4), 2103953. https://doi.org/10.1002/advs.202103953

(7) Forner-Cuenca, A.; Brushett, F. R. Engineering Porous Electrodes for Next-Generation Redox Flow Batteries: Recent Progress and Opportunities. *Curr. Opin. Electrochem.* **2019**, *18*, 113–122. https://doi.org/10.1016/j.coelec.2019.11.002

(8) Warden, C.; Rahman, M. H.; Loud, I. C.; Smith, K. C. Dead Zone and Hot Spot Elimination Using Optimally Tapered Interdigitated Flow Fields: 3D Simulation and Quasi-1D Theory. *J. Electrochem. Soc.* **2025**, *172* (6), 060525. https://doi.org/10.1149/1945-7111/ade3a4

(9) Ke, X.; Prahl, J. M.; Alexander, J. I. D.; Savinell, R. F. Redox Flow Batteries with Serpentine Flow Fields: Distributions of Electrolyte Flow Reactant Penetration into the Porous Carbon Electrodes and Effects on Performance. *J. Power Sources* **2018**, *384*, 295–302. https://doi.org/10.1016/j.jpowsour.2018.03.001

(10) Sauermoser, M.; Kizilova, N.; Pollet, B. G.; Kjelstrup, S. Flow Field Patterns for Proton Exchange Membrane Fuel Cells. *Front. Energy Res.* **2020**, *8* (13), 482555. https://doi.org/10.3389/fenrg.2020.00013

(11) Ke, X.; Prahl, J. M.; Alexander, I. J. D.; Wainright, J. S.; Zawodzinski, T. A.; Savinell, R. F. Rechargeable Redox Flow Batteries: Flow Fields, Stacks and Design Considerations. *Chem. Soc. Rev.* **2018**, *47* (23), 8721–8743. https://doi.org/10.1039/c8cs00072g

(12) Liu, R.; Zhang, Q.; Shen, Y.; Yao, S. Optimization of Flow Field Distribution to Improve Desalination Performance of Battery Electrode Deionization by Designing Flow Channel Structure. *Sep. Purif. Technol.* **2024**, *331*. https://doi.org/10.1016/j.seppur.2023.125661

(13) Saif, H. M.; Gebregeorgis, T. H.; Crespo, J. G.; Pawlowski, S. The Influence of Flow Electrode Channel Design on Flow Capacitive Deionization Performance: Experimental and CFD Modelling Insights. *Desalination* **2024**, *578*, 117452. https://doi.org/10.1016/j.desal.2024.117452

(14) Reale, E. R.; Smith, K. C. Capacitive Performance and Tortuosity of Activated Carbon Electrodes with Macroscopic Pores. *J. Electrochem. Soc.* **2018**, *165* (9), A1685–A1693. https://doi.org/10.1149/2.0601809jes

(15) Huang, Z.; Cai, Y.; Zhang, S.; Wang, R.; Li, X.; Liu, Z. 3D Aligned Architectures for Lithium Batteries: Mechanism, Design, And Manufacture. *Energy Storage Mater.* **2025**, *75*, 103999. https://doi.org/10.1016/j.ensm.2024.103999

(16) Bae, C. J.; Erdonmez, C. K.; Halloran, J. W.; Chiang, Y. M. Design of Battery Electrodes with Dual-Scale Porosity to Minimize Tortuosity and Maximize Performance. *Advanced Materials* **2013**, *25* (9), 1254–1258. https://doi.org/10.1002/adma.201204055

(17) Jeong, H.; Lim, S. J.; Chakravarthy, S.; Kim, K. H.; Lee, J.; Heo, J. S.; Park, H. Three-Dimensional Cathode with Periodically Aligned Microchannels for Improving Volumetric Energy Density of Lithium-Ion Batteries. *J. Power Sources* **2020**, *451*, 227764. https://doi.org/10.1016/j.jpowsour.2020.227764

(18) Li, J.; Luo, S. H.; Ding, X.; Wang, Q.; He, P. NaCl-Template Assisted Synthesis of 3D Honeycomb-Like LiMnPO4/C with High Rate and Stable Performance as Lithium-Ion Battery Cathodes. *ACS Sustain. Chem. Eng.* **2018**, *6* (12), 16683–16691. https://doi.org/10.1021/acssuschemeng.8b03935

(19) Zhang, X.; Hui, Z.; King, S.; Wang, L.; Ju, Z.; Wu, J.; Takeuchi, K. J.; Marschilok, A. C.; West, A. C.; Takeuchi, E. S.; Yu, G. Tunable Porous Electrode Architectures for Enhanced Li-Ion Storage Kinetics in Thick Electrodes. *Nano Lett.* **2021**, *21* (13), 5896–5904. https://doi.org/10.1021/acs.nanolett.1c02142

(20) Yao, C.; Lei, X.; Ma, C.; Li, J.; Liu, J.; Li, C.; Du, P.; Wu, P.; Li, H.; Liu, X. Grooved Cathode Enabled High Areal Capacity Lithium-Air Batteries. *Nano Lett.* **2025**, *25* (20), 8120–8128. https://doi.org/10.1021/acs.nanolett.5c00393

(21) Lu, L. L.; Lu, Y. Y.; Xiao, Z. J.; Zhang, T. W.; Zhou, F.; Ma, T.; Ni, Y.; Yao, H. Bin; Yu, S. H.; Cui, Y. Wood-Inspired High-Performance Ultrathick Bulk Battery Electrodes. *Advanced Materials* **2018**, *30* (20), 1706745. https://doi.org/10.1002/adma.201706745

(22) Wang, X.; Wang, T.; Borovilas, J.; He, X.; Du, S.; Yang, Y. Vertically-Aligned Nanostructures for Electrochemical Energy Storage. *Nano Res.* **2019**, *12* (9), 2002–2017. https://doi.org/10.1007/s12274-019-2392-x

(23) Guyes, E. N.; Simanovski, A.; Suss, M. E. Several Orders of Magnitude Increase in the Hydraulic Permeability of Flow-Through Capacitive Deionization Electrodes via Laser Perforations. *RSC Adv.* **2017**, *7* (34), 21308–21313. https://doi.org/10.1039/c7ra00459a

(24) Bryntesen, S. N.; Finne, P. H.; Svensson, A. M.; Shearing, P. R.; Tolstik, N.; Sorokina, I. T.; Vinje, J.; Lamb, J. J.; Burheim, O. S. Structured Aqueous Processed Lignin-Based NMC Cathodes for Energy-Dense Libs with Improved Rate Capability. *J. Mater. Chem. A Mater.* **2023**, *11* (12), 6483–6502. https://doi.org/10.1039/d2ta08606a

(25) Hille, L.; Noecker, M. P.; Ko, B.; Kriegler, J.; Keilhofer, J.; Stock, S.; Zaeh, M. F. Integration of Laser Structuring into the Electrode Manufacturing Process Chain for Lithium-Ion Batteries. *J. Power Sources* **2023**, *556*, 232478. https://doi.org/10.1016/j.jpowsour.2022.232478

(26) Wang, J.; Sun, Q.; Gao, X.; Wang, C.; Li, W.; Holness, F. B.; Zheng, M.; Li, R.; Price, A. D.; Sun, X.; Sham, T. K.; Sun, X. Toward High Areal Energy and Power Density Electrode for Li-Ion Batteries via Optimized 3D Printing Approach. *ACS Appl. Mater. Interfaces* **2018**, *10* (46), 39794–39801. https://doi.org/10.1021/acsami.8b14797

(27) Park, S.; Shi, B.; Shang, Y.; Deng, K.; Fu, K. Structured Electrode Additive Manufacturing for Lithium-Ion Batteries. *Nano Lett.* **2022**, *22* (23), 9462–9469. https://doi.org/10.1021/acs.nanolett.2c03545

(28) Liu, C.; Qiu, Y.; Liu, Y.; Xu, K.; Zhao, N.; Lao, C.; Shen, J.; Chen, Z. Novel 3D Grid Porous Li4Ti5O12 Thick Electrodes Fabricated by 3D Printing for High Performance

Lithium-Ion Batteries. *Journal of Advanced Ceramics* **2022**, *11* (2), 295–307. https://doi.org/10.1007/s40145-021-0533-7

(29) Yang, Y.; Yuan, W.; Zhang, X.; Yuan, Y.; Wang, C.; Ye, Y.; Huang, Y.; Qiu, Z.; Tang, Y. Overview on the Applications of Three-Dimensional Printing for Rechargeable Lithium-Ion Batteries. *Appl. Energy* **2020**, *257*, 114002. https://doi.org/10.1016/j.apenergy.2019.114002

(30) Wang, H.; Li, J.; Miao, Z.; Huang, K.; Liao, Y.; Xu, X.; Meng, J.; Li, Z.; Huang, Y. Hierarchical Electrode Architecture Enabling Ultrahigh-Capacity LiFePO4 Cathodes with Low Tortuosity. *ACS Appl. Mater. Interfaces* **2023**, *15* (22), 26824–26833. https://doi.org/10.1021/acsami.3c04072

(31) Sandherr, J.; Kleefoot, M. J.; Nester, S.; Weisenberger, C.; DeSilva, A. K. M.; Michel, D.; Reeb, S.; Fingerle, M.; Riegel, H.; Knoblauch, V. Micro Embossing of Graphite-Based Anodes for Lithium-Ion Batteries to Improve Cell Performance. *J. Energy Storage* **2023**, *65*, 107359. https://doi.org/10.1016/j.est.2023.107359

(32) Kim, D.; Magasinski, A.; Lee, S. H.; Yoo, H.; Song, A. Y.; Yushin, G. Micropillar-Based Channel Patterning in High-Loading Graphite Anodes for Superior Li-Ion Batteries. *Battery Energy* **2024**, *3* (3), 20230028. https://doi.org/10.1002/bte2.20230028

(33) Wu, J.; Ju, Z.; Zhang, X.; Quilty, C.; Takeuchi, K. J.; Bock, D. C.; Marschilok, A. C.; Takeuchi, E. S.; Yu, G. Ultrahigh-Capacity and Scalable Architected Battery Electrodes via Tortuosity Modulation. *ACS Nano* **2021**, *15* (12), 19109–19118. https://doi.org/10.1021/acsnano.1c06491

(34) Ren, H.; Wang, Y.; Cao, D.; Gedney, W.; Ji, T.; Sun, X.; Zhu, H. Manufacturing Water-Based Low-Tortuosity Electrodes for Fast-Charge through Pattern Integrated Stamping. *Energy and Environmental Materials* **2023**, *6* (4), e12584. https://doi.org/10.1002/eem2.12584

(35) Johnson, K. E.; Montano, B. C.; Nambu, K. J.; Armstrong, E. N.; Cobb, C. L.; Begley, M. R. Two-Dimensional Patterning of Mesoscale Fibers Using Acoustophoresis. *Mater. Des.* **2023**, *234*, 112328. https://doi.org/10.1016/j.matdes.2023.112328

(36) Zhang, Y.; Shahriar, M.; Hu, S. Structuring Electrodes via Acoustic-Field-Assisted Particle Patterning for Enhanced Performance of Lithium-Ion Batteries. *J. Mater. Chem. A Mater.* **2023**, *11* (22), 11849–11858. https://doi.org/10.1039/d3ta01180a

(37) Pawlowski, S.; Crespo, J. G.; Velizarov, S. Profiled Ion Exchange Membranes: A Comprehensible Review. *Int. J. Mol. Sci.* **2019**, *20* (1), 165. https://doi.org/10.3390/ijms20010165

(38) Vermaas, D. A.; Saakes, M.; Nijmeijer, K. Power Generation Using Profiled Membranes in Reverse Electrodialysis. *J. Memb. Sci.* **2011**, *385–386* (1), 234–242. https://doi.org/10.1016/j.memsci.2011.09.043

(39) Güler, E.; Elizen, R.; Saakes, M.; Nijmeijer, K. Micro-Structured Membranes for Electricity Generation by Reverse Electrodialysis. *J. Memb. Sci.* **2014**, *458*, 136–148. https://doi.org/10.1016/j.memsci.2014.01.060

(40) Liu, J.; Geise, G. M.; Luo, X.; Hou, H.; Zhang, F.; Feng, Y.; Hickner, M. A.; Logan, B. E. Patterned Ion Exchange Membranes for Improved Power Production in Microbial Reverse-Electrodialysis Cells. *J. Power Sources* **2014**, *271*, 437–443. https://doi.org/10.1016/j.jpowsour.2014.08.026

(41) Seo, J.; Kushner, D. I.; Hickner, M. A. 3D Printing of Micropatterned Anion Exchange Membranes. *ACS Appl. Mater. Interfaces* **2016**, *8* (26), 16656–16663. https://doi.org/10.1021/acsami.6b03455

(42) Hasan, M.; Labella, M.; Burke, C. W.; Arges, C. G.; Gomez, E. D.; Gorski, C. A. Dilute Regeneration-Driven Membrane Capacitive Deionization of Synthetic Seawater Using Nanopatterned Membranes and Prussian Blue Analog Electrodes. *Small* **2026**, *22* (3), e10773. https://doi.org/10.1002/smll.202510773

(43) Bhattarai, A.; Wai, N.; Schweiss, R.; Whitehead, A.; Scherer, G. G.; Ghimire, P. C.; Lim, T. M.; Hng, H. H. Vanadium Redox Flow Battery with Slotted Porous Electrodes and Automatic Rebalancing Demonstrated on a 1 kW System Level. *Appl. Energy* **2019**, *236*, 437–443. https://doi.org/10.1016/j.apenergy.2018.12.001

(44) Kumar, R. R.; Suresh, S.; Suthakar, T.; Singh, V. K. Experimental Investigation on PEM Fuel Cell Using Serpentine with Tapered Flow Channels. *Int. J. Hydrogen Energy* **2020**, *45* (31), 15642–15649. https://doi.org/10.1016/j.ijhydene.2020.04.023

(45) Rahman, M. H.; IV, I. C. L.; Do, V. Q.; Hamid, M. A.; Smith, K. C. Tapered, Interdigitated Channels for Uniform, Low-Pressure Flow Through Porous Electrodes for Desalination and Beyond. *Electrochim. Acta* **2025**, *514*, 145632. https://doi.org/10.1016/j.electacta.2024.145632

(46) Beck, V. A.; Ivanovskaya, A. N.; Chandrasekaran, S.; Forien, J. B.; Baker, S. E.; Duoss, E. B.; Worsley, M. A. Inertially Enhanced Mass Transport Using 3D-Printed Porous Flow-Through Electrodes with Periodic Lattice Structures. *PNAS* **2021**, *118* (32), e2025562118. https://doi.org/10.1073/pnas.2025562118

(47) Lisboa, K. M.; Marschewski, J.; Ebejer, N.; Ruch, P.; Cotta, R. M.; Michel, B.; Poulikakos, D. Mass Transport Enhancement in Redox Flow Batteries with Corrugated Fluidic Networks. *J. Power Sources* **2017**, *359*, 322–331. https://doi.org/10.1016/j.jpowsour.2017.05.038

(48) Jangid, M. K.; Davis, A. L.; Liao, D. W.; Dasgupta, N. P. Improved Rate Capability in Composite Solid-State Battery Electrodes Using 3-D Architectures. *ACS Energy Lett.* **2023**, *8* (6), 2522–2531. https://doi.org/10.1021/acsenergylett.3c00528

(49) Dennison, C. R.; Agar, E.; Akuzum, B.; Kumbur, E. C. Enhancing Mass Transport in Redox Flow Batteries by Tailoring Flow Field and Electrode Design. *J. Electrochem. Soc.* **2016**, *163* (1), A5163–A5169. https://doi.org/10.1149/2.0231601jes

(50) Lee, C. H.; Kort-Kamp, W. J. M.; Yu, H.; Cullen, D. A.; Patterson, B. M.; Arman, T. A.; Komini Babu, S.; Mukundan, R.; Borup, R. L.; Spendelow, J. S. Grooved Electrodes for High-Power-Density Fuel Cells. *Nature Energy 2023 8:7* **2023**, *8* (7), 685–694. https://doi.org/10.1038/s41560-023-01263-2

(51) Darling, R. M.; Perry, M. L. The Influence of Electrode and Channel Configurations on Flow Battery Performance. *J. Electrochem. Soc.* **2014**, *161* (9), A1381–A1387. https://doi.org/10.1149/2.0941409jes

(52) Zeng, Y.; Li, F.; Lu, F.; Zhou, X.; Yuan, Y.; Cao, X.; Xiang, B. A Hierarchical Interdigitated Flow Field Design for Scale-Up of High-Performance Redox Flow Batteries. *Appl. Energy* **2019**, *238*, 435–441. https://doi.org/10.1016/j.apenergy.2019.01.107

(53) Lin, T. Y.; Baker, S. E.; Duoss, E. B.; Beck, V. A. Topology Optimization of 3D Flow Fields for Flow Batteries. *J. Electrochem. Soc.* **2022**, *169* (5), 050540. https://doi.org/10.1149/1945-7111/ac716d

(54) Wan, S.; Jiang, H.; Guo, Z.; He, C.; Liang, X.; Djilali, N.; Zhao, T. Machine Learning-Assisted Design of Flow Fields for Redox Flow Batteries. *Energy Environ. Sci.* **2022**, *15* (7), 2874–2888. https://doi.org/10.1039/d1ee03224k

(55) Jiang, K.; Jiang, H.; Zhang, L.; Luan, Y.; Zheng, T.; Liu, M.; Su, X.; Feng, Y.; Lu, G.; Liu, Z. AI-Assisted Design and Optimization of Novel Asymmetric Microchannel Flow Fields for Proton Exchange Membrane Fuel Cells. *Appl. Energy* **2026**, *405*, 127238. https://doi.org/10.1016/j.apenergy.2025.127238

(56) Trogadas, P.; Cho, J. I. S.; Rasha, L.; Lu, X.; Kardjilov, N.; Markötter, H.; Manke, I.; Shearing, P. R.; Brett, D. J. L.; Coppens, M. O. A Nature-Inspired Solution for Water Management in Flow Fields for Electrochemical Devices. *Energy Environ. Sci.* **2024**, *17* (5), 2007–2017. https://doi.org/10.1039/d3ee03666a

(57) Xu, L.; Trogadas, P.; Lan, Y.; Jiang, S.; Zhou, S.; Iacoviello, F.; Du, W.; Jervis, R.; Coppens, M. O. A Nature-Inspired Solution for Water Management in a Zero-Gap CO2 Electrolyzer. *ACS Energy Lett.* **2025**, *10* (7), 3081–3088. https://doi.org/10.1021/acsenergylett.5c01243

(58) Chadwick, E. A.; Shrestha, P.; Parmar, H. B.; Bazylak, A.; Schulz, V. P. Biomimetic Auxiliary Channels Enhance Oxygen Delivery and Water Removal in Polymer Electrolyte Membrane Fuel Cells. *Appl. Energy* **2025**, *389*, 125760. https://doi.org/10.1016/j.apenergy.2025.125760

(59) Zheng, D.; Liang, S.; Zhao, T.; Sun, L.; Fan, W.; Jiang, K.; Cui, H.; Yang, L.; Jia, S.; Lu, G.; Liu, Z. Bio-Inspired Sinusoidally-Waved Flow Fields with Exchange Channels to Enhance PEMFC Performance. *Fuel* **2024**, *367*, 131552. https://doi.org/10.1016/j.fuel.2024.131552

(60) Guo, N.; Leu, M. C.; Koylu, U. O. Bio-Inspired Flow Field Designs for Polymer Electrolyte Membrane Fuel Cells. *Int. J. Hydrogen Energy* **2014**, *39* (36), 21185–21195. https://doi.org/10.1016/j.ijhydene.2014.10.069

(61) Muñoz-Perales, V.; van der Heijden, M.; García-Salaberri, P. A.; Vera, M.; Forner-Cuenca, A. Engineering Lung-Inspired Flow Field Geometries for Electrochemical Flow Cells with Stereolithography 3D Printing. *ACS Sustain. Chem. Eng.* **2023**, *11* (33), 12243–12255. https://doi.org/10.1021/acssuschemeng.3c00848

(62) Trogadas, P.; Cho, J. I. S.; Neville, T. P.; Marquis, J.; Wu, B.; Brett, D. J. L.; Coppens, M. O. A Lung-Inspired Approach to Scalable and Robust Fuel Cell Design. *Energy Environ. Sci.* **2018**, *11* (1), 136–143. https://doi.org/10.1039/c7ee02161e

(63) Kim, Y. Y.; Yang, T. Y.; Suhonen, R.; Välimäki, M.; Maaninen, T.; Kemppainen, A.; Jeon, N. J.; Seo, J. Gravure-Printed Flexible Perovskite Solar Cells: Toward Roll-to-Roll Manufacturing. *Advanced Science* **2019**, *6* (7), 1802094. https://doi.org/10.1002/advs.201802094

(64) Vasconcelos, L. S. de; Xu, R.; Li, J.; Zhao, K. Grid Indentation Analysis of Mechanical Properties of Composite Electrodes in Li-Ion Batteries. *Extreme Mech. Lett.* **2016**, *9*, 495–502. https://doi.org/10.1016/j.eml.2016.03.002

(65) Yao, Q.; Cheng, Y.; Lin, S.; -, al; V Nadimpalli, S. P.; Tripuraneni, R.; Sethuraman -, V. A.; Sheth, J.; Karan, N. K. Stress Evolution in Lithium-Ion Composite Electrodes during Electrochemical Cycling and Resulting Internal Pressures on the Cell Casing. *J. Electrochem. Soc.* **2015**, *162* (14), A2656. https://doi.org/10.1149/2.0341514jes

(66) Smith, K. C.; Mukherjee, P. P.; Fisher, T. S. Columnar Order in Jammed LiFePO4 Cathodes: Ion Transport Catastrophe and Its Mitigation. *Physical Chemistry Chemical Physics* **2012**, *14* (19), 7040–7050. https://doi.org/10.1039/c2cp40135e

(67) Smith, K. C.; Rahman, M. H.; Loud, I. C. I.; Hamid, M. A. Tapered, Interdigitated Channels for Uniform Flow. PCT/US2026/10636, 2026.

(68) Boussinesq. Mémoire Sur l'influence Des Frottements Dans Les Mouvements Réguliers Des Fluides. *J. Math. Pures Appl.* **1868**, *13*, 377–424.

(69) Witherspoon, P. A.; Wang, J. S. Y.; Iwai, K.; Gale, J. E. Validity of Cubic Law for Fluid Flow in a Deformable Rock Fracture. *Water Resour. Res.* **1980**, *16* (6), 1016–1024. https://doi.org/10.1029/wr016I006p01016

(70) Smith, K. C.; Loud, I. C.; Rahman, M. H.; Warden, C.; Do, V. Q. A Compact, Low-Pressure Manifold with Uniform Flow at Low Reynolds Number. *Physics of Fluids* **2024**, *36* (6). https://doi.org/10.1063/5.0211073

(71) Warden, C.; Rahman, M. H.; Loud, I. C.; Smith, K. C. Dead Zone and Hot Spot Elimination Using Optimally Tapered Interdigitated Flow Fields: 3D Simulation and Quasi-1D Theory. *J. Electrochem. Soc.* **2025**, *172*, 060525. https://doi.org/10.1149/1945-7111/ade3a4

(72) Smith, K. C.; Dmello, R. D. Na-Ion Desalination (NID) Enabled by Na-Blocking Membranes and Symmetric Na-Intercalation: Porous-Electrode Modeling. *J. Electrochem. Soc.* **2016**, *163* (3), A530–A539. https://doi.org/10.1149/2.0761603jes

(73) Smith, K. C. Theoretical Evaluation of Electrochemical Cell Architectures Using Cation Intercalation Electrodes for Desalination. *Electrochim. Acta* **2017**, *230*, 333–341. https://doi.org/10.1016/j.electacta.2017.02.006

(74) Liu, S.; Smith, K. C. Quantifying the Trade-Offs between Energy Consumption and Salt Removal Rate in Membrane-Free Cation Intercalation Desalination. *Electrochim. Acta* **2018**, *271*, 652–665. https://doi.org/10.1016/j.electacta.2018.03.065

(75) Liu, S.; Smith, K. C. Quantifying the Trade-Offs between Energy Consumption and Salt Removal Rate in Membrane-Free Cation Intercalation Desalination. *Electrochim. Acta* **2018**, *271*, 652–665. https://doi.org/10.1016/j.electacta.2018.03.065

(76) Shrivastava, A.; Do, V. Q.; Smith, K. C. Efficient, Selective Sodium and Lithium Removal by Faradaic Deionization Using Symmetric Sodium Titanium Vanadium

Phosphate Intercalation Electrodes. *ACS Appl. Mater. Interfaces* **2022**, *14* (27), 30672–30682. https://doi.org/10.1021/acsami.2c03261

(77) Reale, E. R.; Regenwetter, L.; Agrawal, A.; Dardón, B.; Dicola, N.; Sanagala, S.; Smith, K. C. Low Porosity, High Areal-Capacity Prussian Blue Analogue Electrodes Enhance Salt Removal and Thermodynamic Efficiency in Symmetric Faradaic Deionization with Automated Fluid Control. *Water Res. X* **2021**, *13*. https://doi.org/10.1016/j.wroa.2021.100116

(78) Kim, T.; Gorski, C. A.; Logan, B. E. Low Energy Desalination Using Battery Electrode Deionization. *Environ. Sci. Technol. Lett.* **2017**, *4* (10), 444–449. https://doi.org/10.1021/acs.estlett.7b00392

(79) Son, M.; Pothanamkandathil, V.; Yang, W.; Vrouwenvelder, J. S.; Gorski, C. A.; Logan, B. E. Improving the Thermodynamic Energy Efficiency of Battery Electrode Deionization Using Flow-Through Electrodes. *Environ. Sci. Technol.* **2020**, *54* (6), 3628–3635. https://doi.org/10.1021/acs.est.9b06843

(80) Reale, E. R.; Shrivastava, A.; Smith, K. C. Effect of Conductive Additives on the Transport Properties of Porous Flow-through Electrodes with Insulative Particles and Their Optimization for Faradaic Deionization. *Water Res.* **2019**, *165*, 114995. https://doi.org/10.1016/j.watres.2019.114995

(81) Pothanamkandathil, V.; Fortunato, J.; Gorski, C. A. Electrochemical Desalination Using Intercalating Electrode Materials: A Comparison of Energy Demands. *Environ. Sci. Technol.* **2020**, *54* (6), 3653–3662. https://doi.org/10.1021/acs.est.9b07311

(82) Ahn, J.; Kim, S.; Jeon, S. il; Lee, C.; Lee, J.; Yoon, J. Nafion-Coated Prussian Blue Electrodes to Enhance the Stability and Efficiency of Battery Desalination System. *Desalination* **2021**, *500*. https://doi.org/10.1016/j.desal.2020.114778

(83) Beh, E. S.; Benedict, M. A.; Desai, D.; Rivest, J. B. A Redox-Shuttled Electrochemical Method for Energy-Efficient Separation of Salt from Water. *ACS Sustain. Chem. Eng.* **2019**, *7* (15), 13411–13417. https://doi.org/10.1021/acssuschemeng.9b02720

(84) Mohandass, G.; Chen, W.; Krishnan, S.; Kim, T. Asymmetric and Symmetric Redox Flow Batteries for Energy-Efficient, High-Recovery Water Desalination. *Environ. Sci. Technol.* **2022**, *56* (7), 4477–4488. https://doi.org/10.1021/acs.est.1c08609

(85) Do, V. Q.; Smith, K. C. Symmetric Intercalative Desalination of Brackish Water Without Ion-Exchange Membranes. *Environ. Sci. Technol.* **2025**, *59* (31), 16799–16811. https://doi.org/10.1021/acs.est.5c05687

(86) *GitHub - gnea/grbl-Mega: An open source, embedded, high performance g-code-parser and CNC milling controller written in optimized C that will run on an Arduino Mega2560*. https://github.com/gnea/grbl-Mega (accessed 2024-03-21)

(87) Carminati, M.; Luzzatto-Fegiz, P. Conduino: Affordable and High-Resolution Multichannel Water Conductivity Sensor Using Micro USB Connectors. *Sens. Actuators B Chem.* **2017**, *251*, 1034–1041. https://doi.org/10.1016/j.snb.2017.05.184

(88) Zhao, R.; Biesheuvel, P. M.; Miedema, H.; Bruning, H.; van der Wal, A. Charge Efficiency: A Functional Tool to Probe the Double-Layer Structure inside of Porous Electrodes and Application in the Modeling of Capacitive Deionization. *Journal of Physical Chemistry Letters* **2010**, *1* (1), 205–210. https://doi.org/10.1021/jz900154h

# Embossing of Tapered, Isometric Micro-Interdigitated Flow Fields for Enhanced Desalination by Faradaic Deionization

Md Habibur Rahman,[a] Sadia Afrin,[a] and Kyle C. Smith[a,b,c,*]

mhr7@illinois.edu, safrin2@illinois.edu, and kcsmith@illinois.edu

ORCID ID: https://orcid.org/0000-0003-1936-1917, https://orcid.org/0009-0008-6031-4171, and https://orcid.org/0000-0002-1141-1679

[a]Department of Mechanical Science and Engineering, Grainger College of Engineering

[b]Department of Materials Science and Engineering, Grainger College of Engineering

[c]Computational Science and Engineering Program, Grainger College of Engineering

University of Illinois at Urbana-Champaign, Urbana, IL 61801

*Corresponding author's contact information:

Name: Professor Kyle C. Smith

Address: 1206 W. Green St., Urbana, IL 61801

Phone: +1-217-300-0928

Email: kcsmith@illinois.edu

**Table of Contents**

## 1. Nominal Cross-sectional Dimensions of Different Channels

The cross-sectional dimensions of first generation (Gen. I) straight channel and Gen. II piecewise, jointly tapered channel are given in Table S1 and Table S2, respectively.

**Table S1:** Geometric parameters of Gen. I straight channel.

| Parameters | Dimensions |
|---|---|
| Electrode length, $L_e$ $(mm)$ | 45 |
| Channel length, $L_{ch}$ $(mm)$ | 44.027 |
| Channel depth, $h$ $(mm)$ | 0.200 |
| Channel width, $w$ $(mm)$ | 0.262 |
| Flow-path length, $s$ $(mm)$ | 0.638 |
| No. of channels, $N$ $(-)$ | 50 |

**Table S2:** Geometric parameters of segments used to design Gen. II piecewise, jointly tapered channel with embossed, micro-interdigitated flow fields (eμ-IDFFs) in the nickel hexacyanoferrate Prussian blue analog (PBA) electrodes varying both width and depth monotonically along the channel length. Segment 1 corresponds to the first segment at the high-pressure channel inlet, and segment 11 corresponds to the last segment at the channel tip.

| Segment no. | Segment depth (mm) | Segment width (mm) | Segment length (mm) |
|---|---|---|---|
| 1 | 0.200 | 0.350 | 11123 |
| 2 | 0.200 | 0.300 | 11454 |
| 3 | 0.150 | 0.345 | 8084 |
| 4 | 0.150 | 0.290 | 4076 |
| 5 | 0.150 | 0.225 | 3398 |
| 6 | 0.100 | 0.270 | 2159 |
| 7 | 0.100 | 0.215 | 1377 |
| 8 | 0.075 | 0.325 | 937 |
| 9 | 0.075 | 0.275 | 342 |
| 10 | 0.075 | 0.225 | 370 |
| 11 | 0.075 | 0.180 | 707 |

## 2. Channel Characterization

The channel profiles of the embossed PBA electrodes were characterized by optical profiloetry (Keyence VK-X1000). A representative of Gen. I eμ-IDFF is shown in Fig. S1.

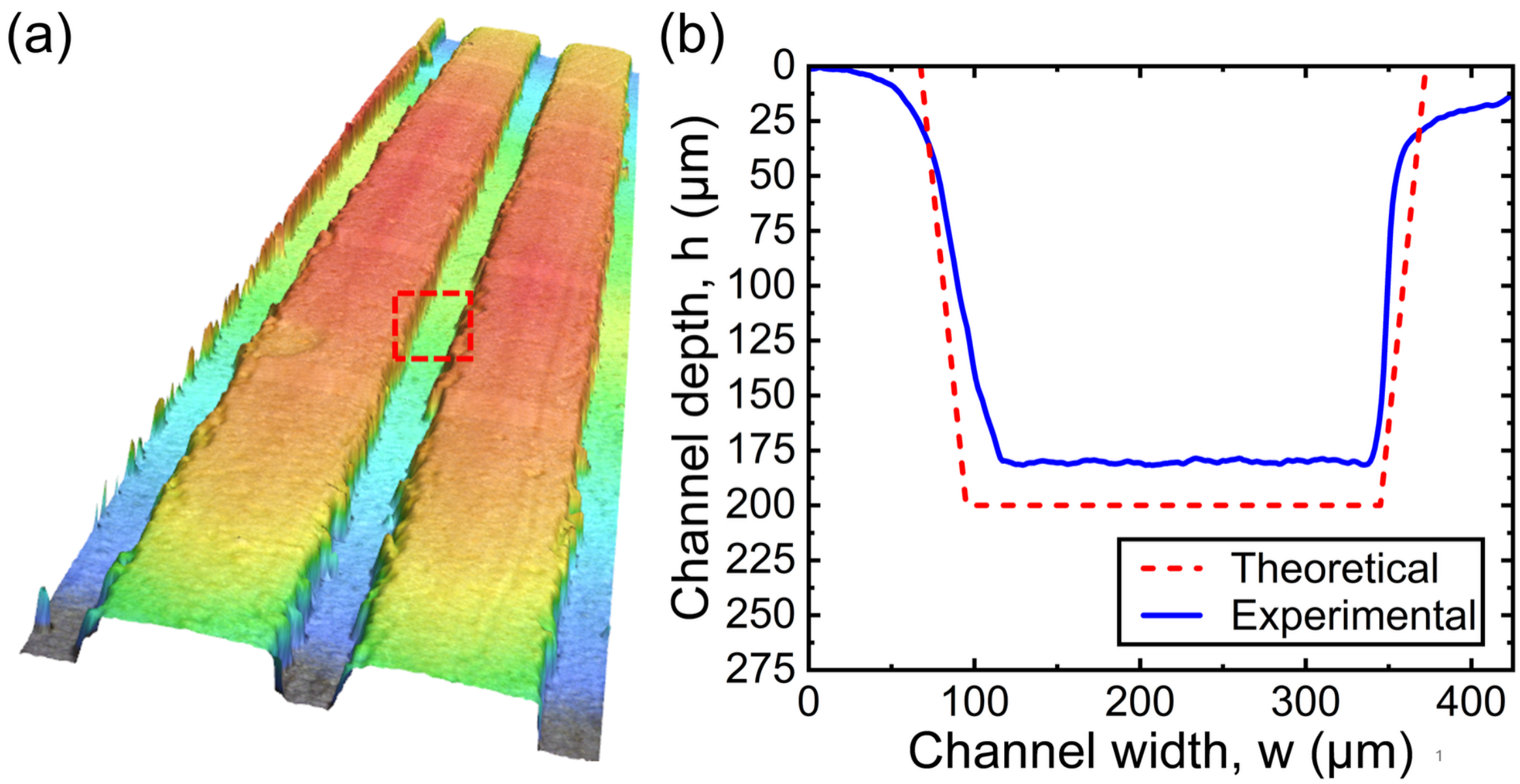

**Figure S1:** (a) Optical profilometry of a representative Gen. I eμ-IDFFs. (b) Measured cross-sectional profile of the corresponding channel, compared with the theoretical channel geometry.

The cross-sectional channel depth profiles of the Gen. III eμ-IDFFs with continuous tapering of channel width were measured by optical profilometry, at different $x$ locations along the channel length, (Fig. S2) and compared with theoretical profiles.

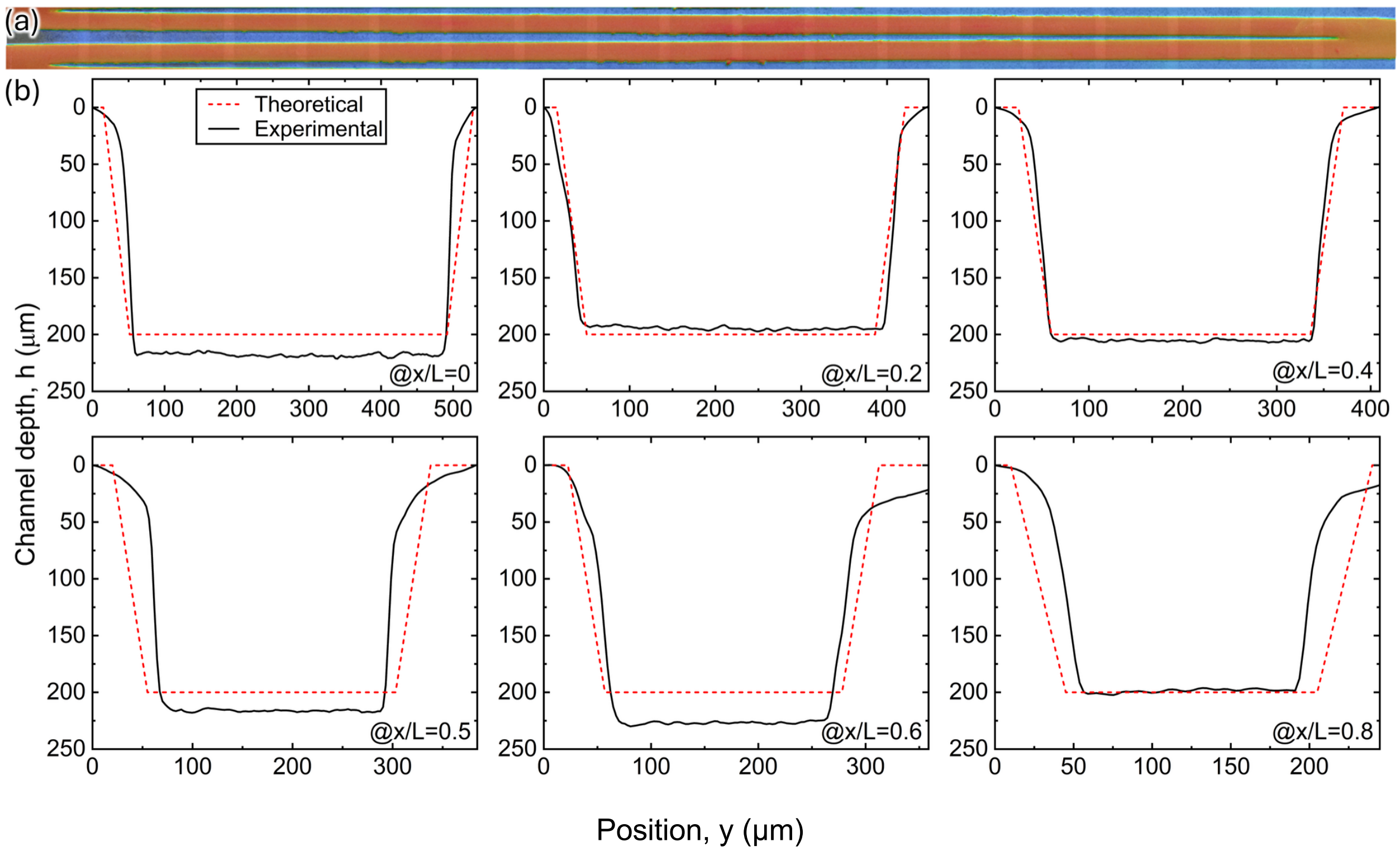

**Figure S2:** (a) Top view of the tapered Gen. III eμ-IDFFs, showing the progressive change in channel width along its length. (b) Cross-sectional channel depth profiles at different normalized positions $(x/L = 0 - 0.8)$ along the channel length, obtained from optical profilometry (solid black lines), and compared with theoretical profiles (dashed red lines). The measured profiles are within ±10% of the nominal dimensions, confirming that the integrated embossing-calendering method successfully fabricates the intended tapered channel with Gen. III eμ-IDFF design.

## 3. Indentation Experiment

A custom-built experimental setup used for the indentation tests, along with a representative indented electrode surface, is shown in Fig. S3.

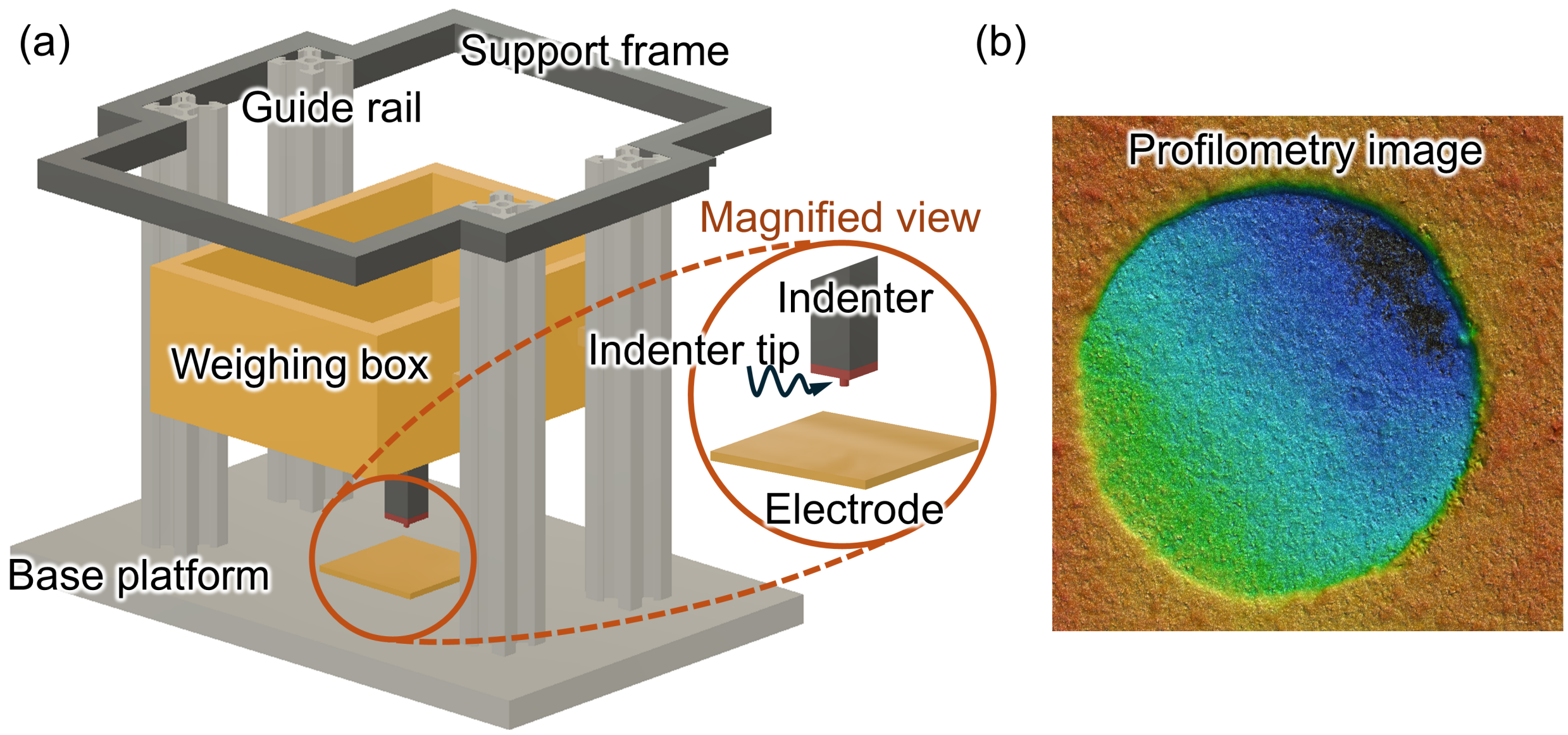

**Figure S3:** (a) In-house built indentation setup showing the guide rails, weighing box, base platform and support frame, with a magnified view of the indenter tip. (b) Optical profilometry of a representative indented electrode surface.

## 4. Quasi-3D Hydrodynamic Model of eμ-IDFFs

### 4.1. Model Overview

The quasi-3D hydrodynamic model determines how the channel geometry controls the uniformity of transverse flow through the porous electrodes patterned with eμ-IDFFs. In this design, fluid supplied to each high-pressure channel is distributed along the channel length, permeates through the intervening porous electrode, and is collected by the low-pressure channels. The transverse flow distribution is therefore governed by the coupling between longitudinal pressure variation within the channels and pressure-driven permeation through the porous electrode.

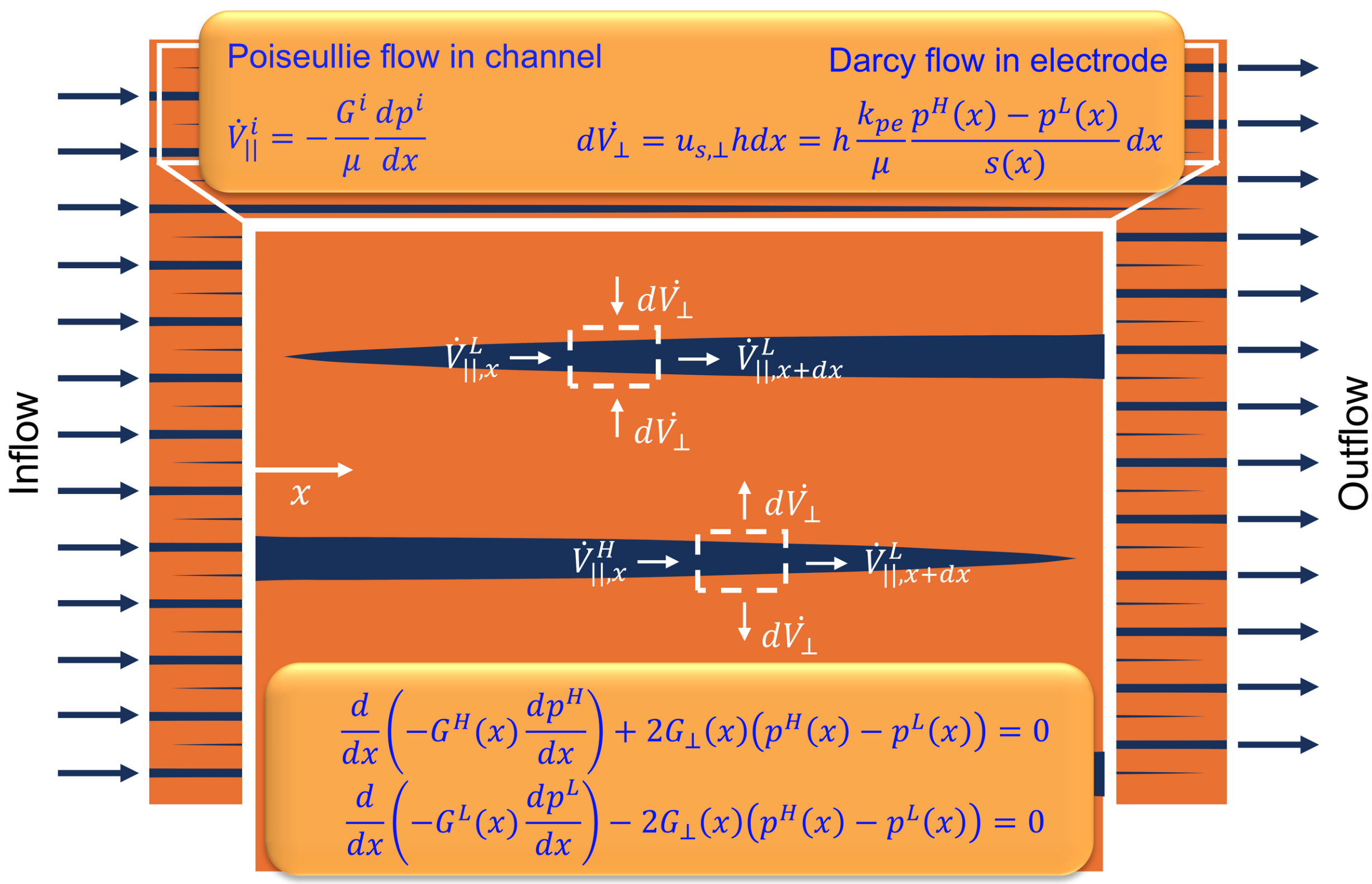

**Figure S4:** Schematic representation of the quasi-3D hydrodynamic model, showing the differential control volume framework used to couple longitudinal pressure variation within interdigitated channels to pressure-driven transverse permeation through the porous electrode.

In this model, pressure and flow rate within each channel are treated as one-dimensional (1D) functions of longitudinal position $x$, whereas flow within the porous electrode is resolved in two dimensions (2D) over the local cross-section spanning the width-wise ($y$) and depth-wise ($z$) directions (Fig. S4 and Fig. S5). The two descriptions are coupled through a transverse conductance $G_{\perp}(x)$, computed from the local 2D flow field at each position $x$. Because the local 2D problem is linear in pressure (SI Sec. 5.1), this $G_{\perp}(x)$ depends only on the local cross-sectional geometry and electrode permeability and is independent of the channel pressures. Throughout, we assume steady, incompressible, laminar flow with negligible inertia through a homogeneous, isotropic porous electrode. We further assume that channel cross-sections vary gradually with $x$, such that flow within each channel is locally fully developed, and that the eμ-IDFF comprises a periodic array of alternating high- and low-pressure channels, such that edge effects are negligible. Because the channel length greatly exceeds the channel spacing $(L \gg s;\ e.g., L = 45\ mm\ and\ s = 643\ \mu m)$, longitudinal flow across the porous electrode is neglected, and transverse flow at each position $x$ is resolved only within the local $y$-$z$ cross-section.

## 4.2. Governing Equations for Longitudinal Channel Flow

At each longitudinal position, the high- and low-pressure channels impose a local inter-channel pressure difference,

$$\Delta p_{\perp}(x) = p^{H}(x) - p^{L}(x)$$

Eq. 1

where $p^{H}(x)$ and $p^{L}(x)$ are the pressures in the high- and low-pressure channels, respectively. This pressure difference is the driving potential for transverse permeation. Because flow between adjacent channels is two-dimensional within the local cross-section, however, it cannot be represented by a single 1D transverse pressure gradient as assumed in our earlier work.[1] Instead, $\Delta p_{\perp}$ is related to the unit-length transverse flow rate $d\dot{V}_{\perp}/dx$ through the transverse conductance $G_{\perp}$:

$$\frac{d\dot{V}_{\perp}}{dx} = \frac{G_{\perp}(x)}{\mu}\Delta p_{\perp}(x)$$

Eq. 2

where $\mu$ is the dynamic viscosity of the working fluid. The transverse conductance, $G_\perp$ represents the integrated hydraulic coupling between the two adjacent channels. It incorporates the effects of channel width, channel depth, electrode permeability, and streamline curvature within the cross-section.

Conservation of mass over a differential control volume of length $dx$ spanning a channel cross-section requires that the longitudinal flow rate in each high-pressure channel decreases as fluid permeates through the porous electrode, and that the longitudinal flow rate in each low-pressure channel increases as the fluid is collected.

This mass balance yields,

$$\frac{d\dot{V}_\parallel^H}{dx} + 2\frac{d\dot{V}_\perp}{dx} = 0$$

Eq. 3

$$\frac{d\dot{V}_\parallel^L}{dx} - 2\frac{d\dot{V}_\perp}{dx} = 0$$

Eq. 4

where $\dot{V}_\parallel^H$ and $\dot{V}_\parallel^L$ are the longitudinal flow rates in the high- and low-pressure channels, respectively. The factor of 2 accounts for the exchange of fluid between each channel and the two neighboring channels on either side.

Assuming locally fully developed Poiseuille flow, the longitudinal flow rate in each channel is expressed in terms of the local longitudinal pressure gradient $dp^i/dx$ through the channel hydraulic conductance defined in the main text:

$$\dot{V}_\parallel^i(x) = -\frac{G^i(x)}{\mu}\frac{dp^i}{dx};\ i = H, L$$

Eq. 5

where $G^i$ is the hydraulic conductance of channel $i$ ($i = H$ for high-pressure channels and $i = L$ for low-pressure channels). Substituting Eq. 2,5 into Eqs. 3,4 yields the coupled pressure equations in the high- and low-pressure channels:

$$\frac{d}{dx}\left[-G^H(x)\frac{dp^H}{dx}\right] + 2G_\perp(x)[p^H(x) - p^L(x)] = 0$$

Eq. 6

$$\frac{d}{dx}\left[-G^L(x)\frac{dp^L}{dx}\right] - 2G_\perp(x)[p^H(x) - p^L(x)] = 0$$

Eq. 7

These two equations form the main quasi-3D hydrodynamic model. Once the pressure profiles $p^H(x)$ and $p^L(x)$ are known, the inter-channel pressure difference, transverse flow rate, transverse superficial velocity, and longitudinal flow rates can be computed directly.

For a channel length $L$, the high-pressure channel is fed at $x = 0$ and closed at $x = L$, whereas the low-pressure channel is closed at $x = 0$ and drained at $x = L$:

$$p^H(0) = p_{in}, \qquad \dot{V}_\parallel^H(L) = 0$$

Eq. 8

$$\dot{V}_\parallel^L(0) = 0, \qquad p^L(L) = p_{out}$$

Eq. 9

Because Eqs. 6-9 are linear, solutions for an arbitrary applied pressure difference are obtained by scaling the solution computed for $p_{in} = 1$ and $p_{out} = 0$.

### 4.3. Longitudinal Hydraulic Conductance of Channels

The longitudinal conductance $G^H(x)$ and $G^L(x)$, which enter the discretized system through the face coefficients of Eq. 22, are determined directly from the local channel geometry. For laminar, fully developed flow through a channel of rectangular cross-section, the conductance of channel $i$ $(H, L)$ is expressed as,

$$G^i(x) = A^i(x)k_{app}^i(x); \qquad A^i(x) = w^i(x)h^i(x)$$

where $A^i(x)$ is the local channel cross-sectional area, and $k_{app}^i$ is the apparent hydraulic permeability, $w^i(x)$ and $h^i(x)$ are the local width and depth of channel i. For a rectangular channel of width $w$ and depth $h$, the apparent permeability follows the exact series solutions of Boussinesq:[2]

$$k_{\text{app}}(w,h) = \frac{w^2}{12}\left[1 - \frac{192w}{\pi^5 h}\sum_{n=1}^{\infty}\frac{\coth(\beta_n h) - \operatorname{csch}(\beta_n h)}{(2n-1)^5}\right]; \qquad \beta_n = \frac{(2n-1)\pi}{w}$$

Eq. 10

The local conductance $G^H(x)$ and $G^L(x)$ are obtained by evaluating Eq. 10 at the local channel width $w^i(x)$ and depth $h^i(x)$. Variations in channel width and depth with $x$ therefore modify $G^H(x)$ and $G^L(x)$ directly, which in turn redistributes the longitudinal pressure profiles in both channels.

### 4.4. Transverse Conductance from Local 2D Darcy Flow

The transverse conductance $G_\perp(x)$ is obtained from the local 2D flow field within the porous electrode. At each position $x_k$, a local cross-section in the $y$-$z$ plane is extracted. It spans from the centerline of a high-pressure channel to the centerline of the adjacent low-pressure channel and contains half of each channel together with the intervening porous electrode (Fig. S5). The local geometry is fully specified by the channel widths $(w^H, w^L)$, channel depths $(h^H, h^L)$, channel center-to-center distance $(s_{cc})$, and electrode thickness $(h)$.

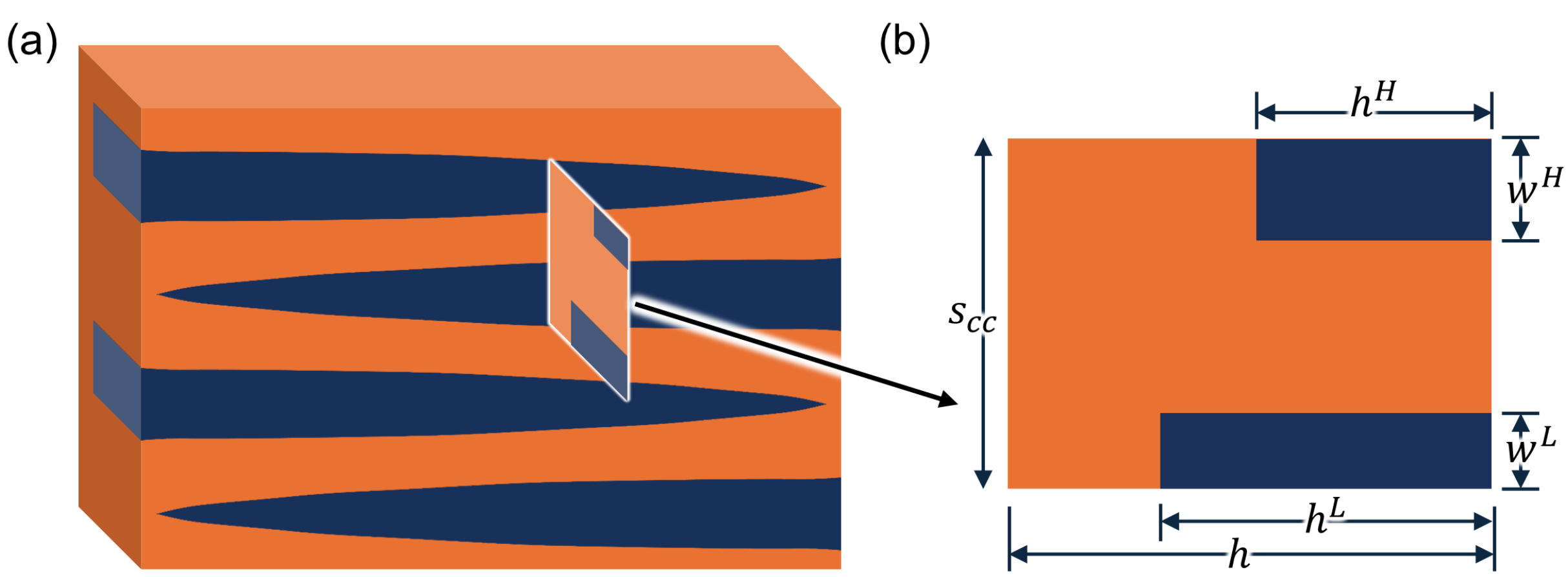

**Figure S5:** Geometry of the local transverse flow domain. (a) Isometric view of alternating high- and low-pressure channels patterned into the porous electrode, with a representative cross-section in the $y$-$z$ plane extracted at position $x$ to define the local

transverse flow domain. (b) Geometric parameters of the local cross-section: electrode thickness $(h)$, channel center-to-center spacing $(s_{cc})$, high- and low-pressure channel widths $(w^H, w^L)$, and corresponding channel depths $(h^H, h^L)$.

Within the porous electrode, the flow is described by Darcy's law:

$$\mathbf{u}_{s,\perp} = -\frac{k_{pe}}{\mu}\nabla p$$

Eq. 11

where $\mathbf{u}_{s,\perp}$ is the local superficial velocity vector and $k_{pe}$ is the electrode permeability. Conservation of mass for an incompressible fluid requires $\nabla \cdot \mathbf{u}_{s,\perp} = 0$. Because $k_{pe}$ is uniform, the pressure field therefore satisfies Laplace's equation:

$$\nabla \cdot \left(\frac{k_{pe}}{\mu}\nabla p\right) = 0, \qquad \frac{\partial^2 p}{\partial y^2} + \frac{\partial^2 p}{\partial z^2} = 0$$

Eq. 12

Pressure within each channel is uniform over the cross-section, consistent with the 1D treatment of channel flow. Dirichlet conditions are imposed on the interfaces between porous electrode and high- and low-pressure channel regions, $p = p^H$ and $p = p^L$, respectively.

No-flux conditions $(\mathbf{n} \cdot \nabla p = 0)$ are applied on all remaining boundaries including channel centerlines, which are planes of symmetry. The transverse flow rate is obtained by integrating the normal component of the superficial velocity across a cut plane $A_\perp$ within the porous electrode that separates channels:

$$\dot{V}_\perp = \int_{A_\perp} \mathbf{u}_{s,\perp} \cdot \mathbf{n}_\perp \, dA$$

Eq. 13

The local transverse conductance then follows from Eq. 2:

$$G_\perp(x) = \frac{\mu}{p^H(x) - p^L(x)} \frac{d\dot{V}_\perp}{dx}(x)$$

Eq. 14

Because Eq. 12 is linear and the pressures enter only through the boundary conditions, $G_\perp(x)$ is independent of the applied pressures. Each cross-section is therefore solved once with $p^H = 1$ and $p^L = 0$. In the limiting case of full-depth channels with straight walls and unidirectional transverse flow across flow path length $s(x)$, Eq. 14 reduces to,

$$G_\perp(x) = \frac{k_{pe} h}{s(x)}$$

Eq. 15

Deviations from Eq. 15 therefore quantify the influence of streamline curvature caused by incomplete channel penetration and by variations in channel geometry. For Gen. III eμ-IDFFs, the isometric constraint maintains a constant flow-path $s$ across the porous electrode between adjacent high- and low-pressure channels along $x$, which explains the nearly constant $G_\perp$ (Fig. 2b).

## 5. Numerical Discretization

### 5.1. Finite-Volume Discretization of the Local Transverse Pressure Field

The local 2D pressure equation (Eq. 12) is discretized using a cell-centered finite-volume method. Integrating $\nabla \cdot u_{s,\perp} = 0$ over a control volume and applying the divergence theorem gives

$$\int_{CS} \left(-\frac{k_{pe}}{\mu}\nabla p\right) \cdot \mathbf{n}\, dA = 0$$

Eq. 16

where $CS$ is the control surface and $\mathbf{n}$ is the outward unit normal vector. We consider a 2D-control volume centered at node $P$, with neighboring nodes $E$ and $W$ in the $\pm y$ directions, and $N$ and $S$ in the $\pm z$ directions. The control volume has dimensions $\Delta y$ and $\Delta z$, the distances from node $P$ to its neighbors are $\delta y_e$, $\delta y_w$, $\delta z_n$ and $\delta z_s$, and all quantities are expressed per unit length in the $x$ direction. The flux balance over the four faces of the control volume gives,

$$\int_s^n \left[\left(\frac{k_{pe}}{\mu}\frac{\partial p}{\partial y}\right)_e - \left(\frac{k_{pe}}{\mu}\frac{\partial p}{\partial y}\right)_w\right] dz + \int_w^e \left[\left(\frac{k_{pe}}{\mu}\frac{\partial p}{\partial z}\right)_n - \left(\frac{k_{pe}}{\mu}\frac{\partial p}{\partial z}\right)_s\right] dy = 0$$

Approximating each face integral by its face-center value:

$$\left[\left(\frac{k_{pe}}{\mu}\frac{\partial p}{\partial y}\right)_e - \left(\frac{k_{pe}}{\mu}\frac{\partial p}{\partial y}\right)_w\right] \Delta z + \left[\left(\frac{k_{pe}}{\mu}\frac{\partial p}{\partial z}\right)_n - \left(\frac{k_{pe}}{\mu}\frac{\partial p}{\partial z}\right)_s\right] \Delta y = 0$$

The face-normal pressure gradients are approximated using central differences:

$$\left(\frac{\partial p}{\partial y}\right)_e = \frac{p_E - p_P}{\delta y_e}, \qquad \left(\frac{\partial p}{\partial y}\right)_w = \frac{p_P - p_W}{\delta y_w}$$

$$\left(\frac{\partial p}{\partial z}\right)_n = \frac{p_N - p_P}{\delta z_n}, \qquad \left(\frac{\partial p}{\partial z}\right)_s = \frac{p_P - p_S}{\delta z_s}$$

Substituting these approximations gives

$$\frac{k_{pe}}{\mu}\left[\frac{p_E - p_P}{\delta y_e} - \frac{p_P - p_W}{\delta y_w}\right] \Delta z + \frac{k_{pe}}{\mu}\left[\frac{p_N - p_P}{\delta z_n} - \frac{p_P - p_S}{\delta z_s}\right] \Delta y = 0$$

This can be rearranged into the standard finite-volume coefficient form,

$$a_P p_P = a_E p_E + a_W p_W + a_N p_N + a_S p_S + b_p$$

Eq. 17

where

$$a_E = \frac{k_{pe}}{\mu}\frac{\Delta z}{\delta y_e};\ a_W = \frac{k_{pe}}{\mu}\frac{\Delta z}{\delta y_w};\ a_N = \frac{k_{pe}}{\mu}\frac{\Delta y}{\delta z_n};\ a_S = \frac{k_{pe}}{\mu}\frac{\Delta y}{\delta z_s}$$

and $a_P = a_E + a_W + a_N + a_S\ and\ b_p$ is nonzero only adjacent to prescribed-pressure boundaries. On a uniform rectangular grid, the coefficients reduce to,

$$a_y = a_E = a_W = \frac{k_{pe}}{\mu}\frac{\Delta z}{\Delta y}, \qquad a_z = a_N = a_S = \frac{k_{pe}}{\mu}\frac{\Delta y}{\Delta z}$$

Eq. 18

Here, $a_y$ and $a_z$ correspond to fluxes through the east-west and north-south faces, respectively, which are generally unequal because the grid spacing may differ between directions ($\Delta y \neq \Delta z$). For an internal porous cell, Eq. 17 becomes,

$$(2a_y + 2a_z)p_P - a_y p_E - a_y p_W - a_z p_N - a_z p_S = 0$$

Eq. 19

*No-flux boundaries:* No-flux conditions are applied at all external solid boundaries by omitting the flux contribution through the corresponding face from the finite-volume balance. For example, at a west no-flux boundary ($\partial p/\partial y = 0$), the west-neighbor term $a_W p_W$ is excluded and $a_W$ is excluded from $a_P$ entirely.

*Prescribed pressure boundaries:* Channel pressure is imposed at the interface between a porous cell and the adjacent channel region. Because this interface lies one half-cell from the porous-cell center, the node-to-boundary distance is halved, and the corresponding coefficient is doubled:

$$a_y^* = 2a_y$$
$$a_z^* = 2a_z$$

Eq. 20

Channel cells are retained in the vector of unknowns and are assigned identity rows in the coefficient matrix, which fix their values to the prescribed channel pressure.

*Illustrative Example:* To illustrate the assembly procedure, we consider a representative $3 \times 3$ grid with cells numbered column-wise, such that columns are arranged along $y$ and rows along $z$,

$$\begin{matrix} p_3 & p_6 & p_9 \\ p_2 & p_5 & p_8 \\ p_1 & p_4 & p_7 \end{matrix}$$

Cells 4 and 7 lie within the low-pressure channel and cell 9 lies within the high-pressure channel, such that $p_4 = p_7 = p^L$ and $p_9 = p^H$. All external boundaries of the grid are no-flux boundaries. The prescribed-pressure channel cells are imposed directly using identity rows in the coefficient matrix:

$$A_{4,4} = 1,\ b_4 = p^L$$

$$A_{7,7} = 1,\ b_7 = p^L$$

$$A_{9,9} = 1,\ b_9 = p^H$$

The resulting linear system $A\boldsymbol{p} = \boldsymbol{b}$ with pressure vector, $\boldsymbol{p} = [p_1, p_2, p_3, \dots, p_9]^T$ is,

$$A = \begin{bmatrix} 2a_y + a_z & -a_z & 0 & 0 & 0 & 0 & 0 & 0 & 0 \\ -a_z & a_y + 2a_z & -a_z & 0 & -a_y & 0 & 0 & 0 & 0 \\ 0 & -a_z & a_y + a_z & 0 & 0 & -a_y & 0 & 0 & 0 \\ 0 & 0 & 0 & 1 & 0 & 0 & 0 & 0 & 0 \\ 0 & -a_y & 0 & 0 & 2a_y + 3a_z & -a_z & 0 & -a_y & 0 \\ 0 & 0 & -a_y & 0 & -a_z & 3a_y + a_z & 0 & 0 & 0 \\ 0 & 0 & 0 & 0 & 0 & 0 & 1 & 0 & 0 \\ 0 & 0 & 0 & 0 & -a_y & 0 & 0 & a_y + 4a_z & 0 \\ 0 & 0 & 0 & 0 & 0 & 0 & 0 & 0 & 1 \end{bmatrix}$$

and

$$\boldsymbol{b} = \begin{bmatrix} 2a_y p^L \\ 0 \\ 0 \\ p^L \\ 2a_z p^L \\ 2a_y p^H \\ p^L \\ 2a_z p^L + 2a_z p^H \\ p^H \end{bmatrix}$$

Eq. 21

The example illustrates the three types of cells handled by the local solver. Prescribed pressure channel cells (cell 4, 7 and 9) are imposed through identity rows, and their pressures enter adjacent porous cells as source terms in $b$. Porous cells exchange fluids with porous neighbors through the standard coefficients of Eq. 19, with two modifications. First, faces on no-flux boundaries are omitted, which reduces the diagonal coefficient (e.g., cells 1, 2, and 3). Second, faces adjacent to channel cells use the doubled coefficients of Eq. 20. For example, cell 5 exchanges fluid with porous cells 2, 6, and 8 through standard coefficients, but its south face borders low-pressure channel cell 4, such that $a_P = a_y^* + a_z = 2a_y + 3a_z$ and $b_5 = 2a_z p^L$.

After the local pressure field is solved, the discrete form of Eq. 14 is evaluated by summing the fluxes through all faces on the high-pressure interface.

### 5.2. Finite-Volume Discretization of the Coupled Channel Equations

The channel length $L$ is divided into $N$ uniform cells of length $\Delta x = L/N$ with centers at $x_k = \left(k - \frac{1}{2}\right)\Delta x$. Integrating Eqs. 6 and 7 over cell $k$ and approximating face fluxes by central differences yields equations expressed in terms of the face conductance coefficients,

$$H_k^{\pm} = \frac{G^H_{k\pm\frac{1}{2}}}{\Delta x}, \qquad L_k^{\pm} = \frac{G^L_{k\pm\frac{1}{2}}}{\Delta x}$$

Eq. 22

where $G^i_{k\pm\frac{1}{2}}$ is evaluated at the cell faces $x_{k\pm\frac{1}{2}}$.

The transverse exchange coefficient is then introduced as

$$\lambda_k = 2G_{\perp,k}\Delta x$$

Eq. 23

where $G_{\perp,k} = G_\perp(x_k)$ and the factor of 2 follows from Eq. 3 and Eq. 4. For interior cells $(2 \le k \le N-1)$, the discretized equations are,

$$-H_k^- p_{k-1}^H + (H_k^- + H_k^+ + \lambda_k) p_k^H - H_k^+ p_{k+1}^H - \lambda_k p_k^L = 0$$

Eq. 24

$$-\lambda_k p_k^H - L_k^- p_{k-1}^L + (L_k^- + L_k^+ + \lambda_k) p_k^L - L_k^+ p_{k+1}^L = 0$$

Eq. 25

Each high-pressure cell is therefore coupled only to its neighboring high-pressure cells and to the low-pressure cell at the same position, and vice versa. Boundary conditions (Eqs. 8 and 9) are imposed in the first and last cells. Closed channel ends are treated by omitting the corresponding face flux. Prescribed pressures are imposed at the boundary face, one half-cell from the adjacent cell center, such that the corresponding coefficient is doubled. The resulting boundary equations are,

High-pressure inlet $(k = 1)$: $(2H_1^- + H_1^+ + \lambda_1) p_1^H - H_1^+ p_2^H - \lambda_1 p_1^L = 2H_1^- p_{in}$

High-pressure closed tip $(k = N)$: $-H_N^- p_{N-1}^H + (H_N^- + \lambda_N) p_N^H - \lambda_N p_N^L = 0$

Low-pressure closed tip $(k = 1)$: $-\lambda_1 p_1^H + (L_1^+ + \lambda_1) p_1^L - L_1^+ p_2^L = 0$

Low-pressure outlet $(k = N)$: $-\lambda_N p_N^H - L_N^- p_{N-1}^L + (L_N^- + 2L_N^+ + \lambda_N) p_N^L = 2L_N^+ p_{out}$

Eq. 26

Arranging the unknowns as $\boldsymbol{p} = [p_1^H \quad p_2^H \quad \cdots \quad p_N^H \quad p_1^L \quad p_2^L \quad \cdots \quad p_N^L]^T$, yields the sparse linear system,

$$A\boldsymbol{p} = \boldsymbol{b}$$

Eq. 27

Longitudinal channel flow produces tridiagonal diagonal blocks of $A$, while transverse permeation produces diagonal off-diagonal blocks. Because $H_k^+ = H_{k+1}^-$ and $L_k^+ = L_{k+1}^-$, $A$ is symmetric. For example, for $N = 3$, the unknown pressure vector is $\boldsymbol{p} = [p_1^H, p_2^H, p_3^H, p_1^L, p_2^L, p_3^L]$, and the coupled sparse matrix is

$$A = \begin{bmatrix} 2H_1^- + H_1^+ + \lambda_1 & -H_1^+ & 0 & -\lambda_1 & 0 & 0 \\ -H_2^- & H_2^- + H_2^+ + \lambda_2 & -H_2^+ & 0 & -\lambda_2 & 0 \\ 0 & -H_3^- & H_3^- + \lambda_3 & 0 & 0 & -\lambda_3 \\ -\lambda_1 & 0 & 0 & L_1^+ + \lambda_1 & -L_1^+ & 0 \\ 0 & -\lambda_2 & 0 & -L_2^- & L_2^- + L_2^+ + \lambda_2 & -L_2^+ \\ 0 & 0 & -\lambda_3 & 0 & -L_3^- & L_3^- + 2L_3^+ + \lambda_3 \end{bmatrix}$$

and

$$\boldsymbol{b} = \begin{bmatrix} 2H_1^- p_{in} \\ 0 \\ 0 \\ 0 \\ 0 \\ 2L_3^+ p_{out} \end{bmatrix}$$

Eq. 28

For the unit-pressure solution, $p_{in} = 1$ and $p_{out} = 0$, only the first entry of $b$ is nonzero. The model was implemented in MATLAB, and the sparse linear systems for the local transverse pressure field (Eqs. 17-20) and the coupled channel equations (Eq. 27) were solved using MATLAB's direct sparse solver.

### 5.3. Post-Processing

The solution vector is separated into the channel pressure profiles, $p_k^H = P_k,\ p_k^L = P_{N+k}$, $k = 1,2, \dots, N$. The local inter-channel pressure difference and the unit-length transverse flow rate are,

$$\Delta p_{\perp,k} = p_k^H - p_k^L, \qquad \frac{d\dot{V}_{\perp,k}}{dx} = \frac{G_{\perp,k}}{\mu}(p_k^H - p_k^L)$$

Eq. 29

The transverse superficial velocity averaged over the electrode thickness follows as

$$\langle u_{s,\perp} \rangle_k = \frac{1}{h}\frac{d\dot{V}_{\perp,k}}{dx} = \frac{G_{\perp,k}}{\mu h}(p_k^H - p_k^L)$$

Eq. 30

and the longitudinal flow rates at cell faces are recovered from Eq. 5:

$$\dot{V}^H_{||,k+\frac{1}{2}} = -\frac{G^H_{k+\frac{1}{2}}}{\mu}\frac{p^H_{k+1} - p^H_k}{\Delta x}, \qquad \dot{V}^L_{||,k+\frac{1}{2}} = -\frac{G^L_{k+\frac{1}{2}}}{\mu}\frac{p^L_{k+1} - p^L_k}{\Delta x}$$

Eq. 31

For fixed $\mu$ and $h$, uniform transverse velocity requires that the product $G_\perp(x)[p^H(x) - p^L(x)]$ be constant along $x$. Modifying channel profiles therefore improves flow uniformity through two simultaneous mechanisms: redistributing the longitudinal pressure profiles and modifying the local transverse conductance. We compare Gen. I, Gen. II, and Gen. III eμ-IDFFs on this basis by examining their channel profiles, pressure distributions, transverse conductance, transverse superficial velocity, and representative 2D streamline contours, as shown in Fig. 2.

## 6. Closed-Form Mechanical Model for Roll Embossing of Porous Electrodes

We determine a linear relationship between the average normal pressure and the local electrode deformation from indentation experiments (Fig. 1c), expressed as

$$\langle p \rangle = \kappa \left( h_{x=x_i} - h(x) \right) = \kappa \left( h_i - h(x) \right)$$

Eq. 1

where $\langle p \rangle$ is the average normal pressure, $h_i$ is the thickness of the electrode strip (unit in m) before undergoing any deformation, $h(x)$ is the local thickness along the longitudinal direction within the roll gap. We obtain a proportionality constant $\kappa = 36\ MPa/m$ from the indentation measurements.

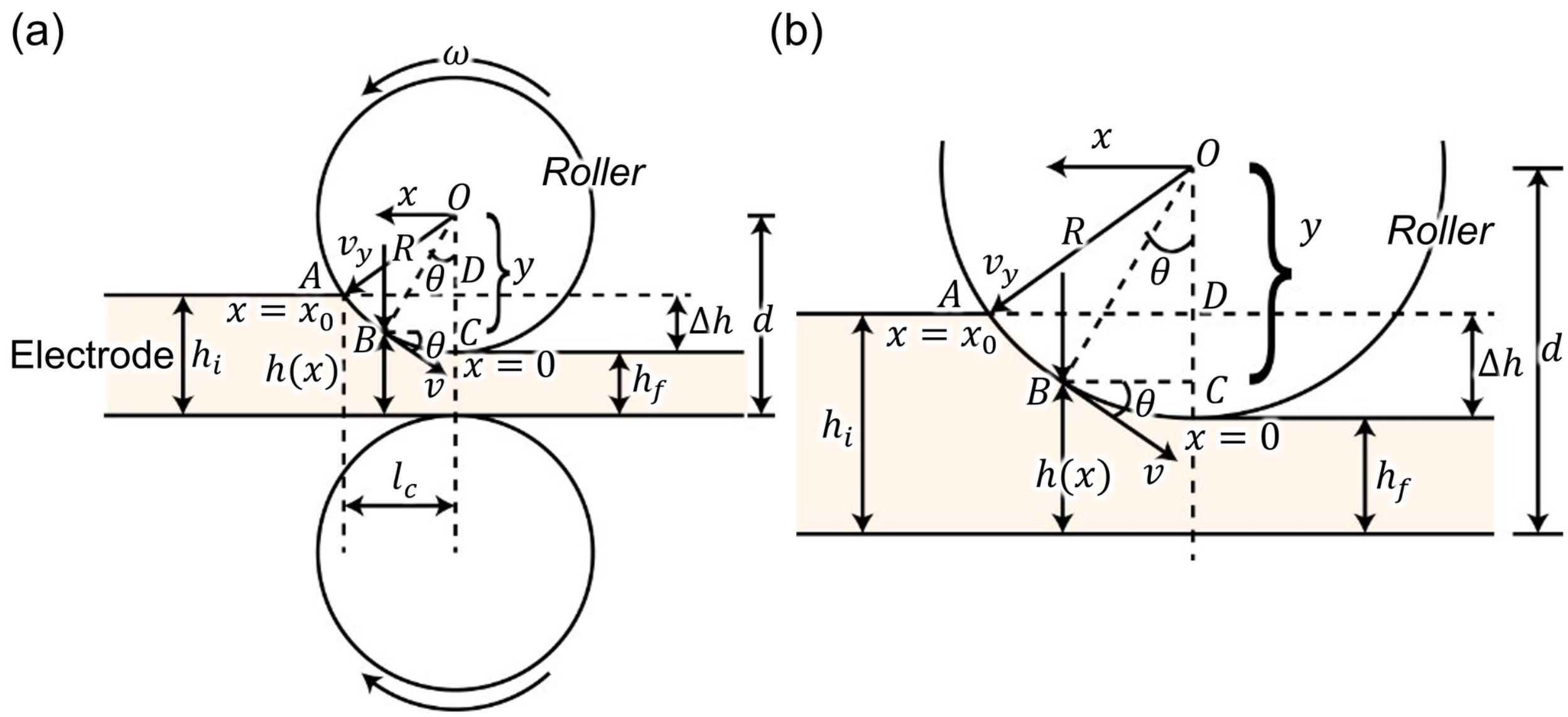

**Figure S6:** (a) Schematic of the electrode passing through the rollers during integrated embossing-calendering method. (b) Enlarged view of the roll-electrode contact region, showing the plastically deformed electrode geometry within the roll gap.

We describe the deformation geometry by defining the distance from the top roller center to the electrode base as $d = R + h_f(x = 0) = y + h(x)$, where $R$ is the roller radius, $h_f$ is the deformed electrode thickness at the exit. Using this geometric relation, we express the local thickness as $h(x) = d - \sqrt{R^2 - x^2}$. The contact length $x_c$ between the roller and the electrode can be determined from triangle OAD using $R^2 = x_c^2 + (R - \Delta h)^2$, where $\Delta h = h_i - h_f$. Assuming $\Delta h \ll R$ we can simplify this expression to $x_c = \sqrt{2R\Delta h}$.

We then compute the rolling force per unit width by integrating the pressure distribution over the contact length,

$$F_R' = \int_0^{x_c} \langle p \rangle dx = \int_0^{x_c} \kappa \left( h_i - d + \sqrt{R^2 - x^2} \right) dx$$

Eq. 2

which yields

$$F_R' = \kappa \left[ \frac{x_c}{2} \sqrt{R^2 - x_c^2} + \frac{R^2}{2} \sin^{-1}\left(\frac{x_c}{R}\right) + (h_i - d) x_c \right]$$

Eq. 3

For $x_c \ll R$, we expand the arcsine and square-root terms to third order in $x_c/R$, $\sin^{-1}(x_c/R) \approx x_c/R + (1/6)(x_c/R)^3$ and $\sqrt{R^2 - x_c^2} \approx R - x_c^2/(2R)$ and substitute $d = R + h_f$, which gives $x^3$

$$F_R' = \kappa \left[ x_c \Delta h - \frac{x_c^3}{6R} \right] = \frac{2}{3} \kappa x_c \Delta h = \frac{2}{3} \kappa \Delta h \sqrt{2R\Delta h}$$

Eq. 4

The cubic term cannot be neglected even though $x_c \ll R$, because the term $x_c^3/(6R) = x_c \Delta h/3$ is the same order as the leading term $x_c \Delta h$.

We now apply an energy balance to the control volume defined by the rollers and the deforming electrode strip,

$$\dot{E}_{in} - \dot{E}_{out} + \dot{E}_g = \dot{E}_{st} + \dot{W}_{in}$$

Eq. 5

where $\dot{E}_{in}$ and $\dot{E}_{out}$ represent the energy entering and leaving the control volume with the electrode strip, $\dot{E}_g$ represents the energy dissipated due to plastic deformation, $\dot{E}_{st}$ is the change of stored energy in the system and $\dot{W}_{in}$ is the mechanical work input from the rollers to overcome deformation resistance and drive the electrode strip through the roll gap. We consider equilibrium steady-state conditions such that $\dot{E}_{in} = \dot{E}_{out}$ and $\dot{E}_{st} = 0$ which simplifies the energy balance to $\dot{E}_g = \dot{W}_{in}$.

We can write the energy dissipated due to plastic deformation as $\dot{E}_g = \int_0^{l_c} \langle p \rangle v_y z dx$ where $z$ is the electrode width and $v_y$ is the vertical component of the tangential velocity. Using the roller kinematics $v_y = v \sin\theta = \omega R.x/R = \omega x$, we can rewrite the expression as

$$\dot{E}_g = \int_0^{x_c} \kappa \left( h_i - d + \sqrt{R^2 - x^2} \right) \omega z x dx$$

We express the mechanical power input as $\dot{W}_{in} = T_R \omega$ where $T_R$ is the rolling torque and obtain the rolling torque per unit width as

$$T_R' = \int_0^{x_c} \kappa \left( h_i - d + \sqrt{R^2 - x^2} \right) x dx$$

which yields

$$T_R' = \kappa \left[ (h_i - d) \frac{x_c^2}{2} - \frac{1}{3} (R^2 - x_c^2)^{3/2} + \frac{1}{3} R^3 \right]$$

For $x_c \ll R$, we expand the $(R^2 - x_c^2)^{3/2}$ term to fourth order in $x_c$, $(R^2 - x_c^2)^{3/2} \approx R^3 - 3R x_c^2/2 + 3x_c^4/(8R)$, and substitute $d = R + h_f$, which gives

$$T_R' = \kappa \left[ \frac{\Delta h x_c^2}{2} - \frac{x_c^4}{8R} \right] = \frac{1}{4} \kappa x_c^2 \Delta h = \frac{1}{4} k \Delta h \sqrt{2R\Delta h}$$

Eq. 6

Finally, we can obtain the effective arm using $T_R' = r_{eff} F_R'$ as

$$r_{eff} = \frac{T_R'}{F_R'} = \frac{3}{8} x_c = \frac{3}{8} \sqrt{2R\Delta h}$$

We now consider electrodes patterned with Gen. III eµ-IDFFs having isometric flow paths through the porous electrodes and assume negligible gaps at both the channel ends and the electrode side boundaries. At any given location along the channel length, we write the total rolling force as

$$F_{R,t} = N_H w_H(x) F'_{R_H} + N_L w_L(x) F'_{R_L} + (N-1) s(x) F'_{R_{s(x)}}$$

Eq. 7

where $w(x)$ is the channel width at position $x$, $N$ is the total number of channels, $s(x)$ is the inter-channel distance as a function of $x$. Subscripts $H$ and $L$ denote high- and low-pressure channels, respectively.

For an electrode that has no channels and is simply compressed to produce a certain thickness change $\Delta h$, the stamping force $F'_S$ per unit electrode width is shown to scale linearly with $\Delta h$:

$$F'_S = \kappa L \Delta h$$

where $L$ is overall electrode length. We can also calculate the stamping force required to pattern high- and low-pressure channels by considering an infinitesimal segment of $dx$ along the electrode's length. Here, the differential stamping force acting on that segment is:

$$dF_S(x) = p_H dA_H(x) + p_L dA_L(x) + p_s dA_s(x)$$

Eq. 8

where $p_H$, $p_L$, and $p_s$ are the pressures applied on the high-pressure channel, the low-pressure channel and in the bulk electrode region, respectively. For an electrode we define the corresponding areas within the segments as $dA_H(x) = N_H w_H(x) dx$, $dA_L(x) = N_L w_L(x) dx$ and $dA_s(x) = [W - N_H w_H(x) - N_L w_L(x)] dx$, where $W$ is the electrode width. We then substitute these expressions in Eq. 8 to have the local stamping force per unit length as-

$$F'_S(x) = \frac{dF_S(x)}{dx} = N_H w_H(x)(p_H - p_s) + N_L w_L(x)(p_L - p_s) + p_s W$$

We can simplify this expression by considering $N_H = N_L = N/2$, $p_H = p_L = p_c$ and $w_H(x) + w_L(x) = w_t =$ constant:

$$F'_S(x) = \frac{N}{2} w_t (p_c - p_s) + W p_s = constant$$

We can obtain the total stamping force by integrating this expression along the electrode length $L$:

$$F_{S,t} = \int_0^L F_S'(x)dx = L\left[\frac{N}{2}w_t(p_c - p_s) + Wp_s\right]$$

Eq. 9

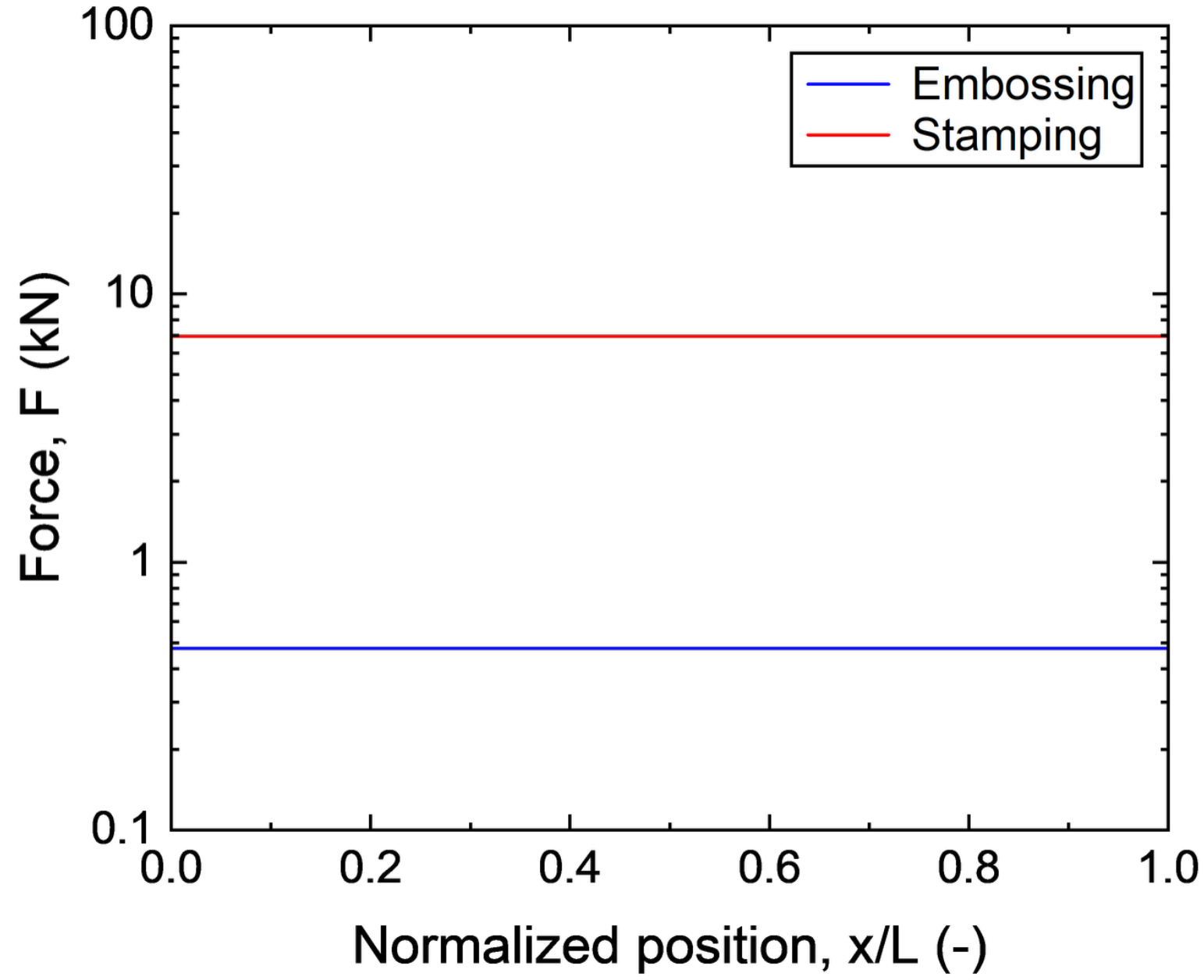

**Figure S7:** Force required to pattern channels into electrode with Gen. III eμ-IDFFs using both embossing and stamping techniques.

We finally apply the model to a Gen. III eμ-IDFF-embossed electrode (Fig. 1b) where each channel has a length 44.027 mm, a maximal width of 514 μm, and a depth of 200 μm. The center-to-center spacing between two adjacent channels is 900 μm and the electrode contains 25 high-pressure and 25 low-pressure channels. Using the rolling-force (Eq. 7) and stamping-force (Eq. 9) expressions, we calculate the total force required to emboss and stamp the channel patterns into the electrode as a function of position. The results show that the embossing method reduces the required force by 15-fold relative to conventional stamping (Fig. S7).

## 7. X-ray Computed Tomography (CT) Scanner and Reconstructed 3D Images

To understand the effect of embossing on the internal microstructure of the porous electrodes, X-ray CT was conducted at 20x magnification over an electrode area of 0.548 $\mu m^2$. Three Gen. III electrodes embossed with different die heights of 50, 200, and 400 µm were examined to evaluate the influence fo embossing depth (Fig. S8). The scan settings for the samples with different degrees of embossing depth are summarized in Table S3.

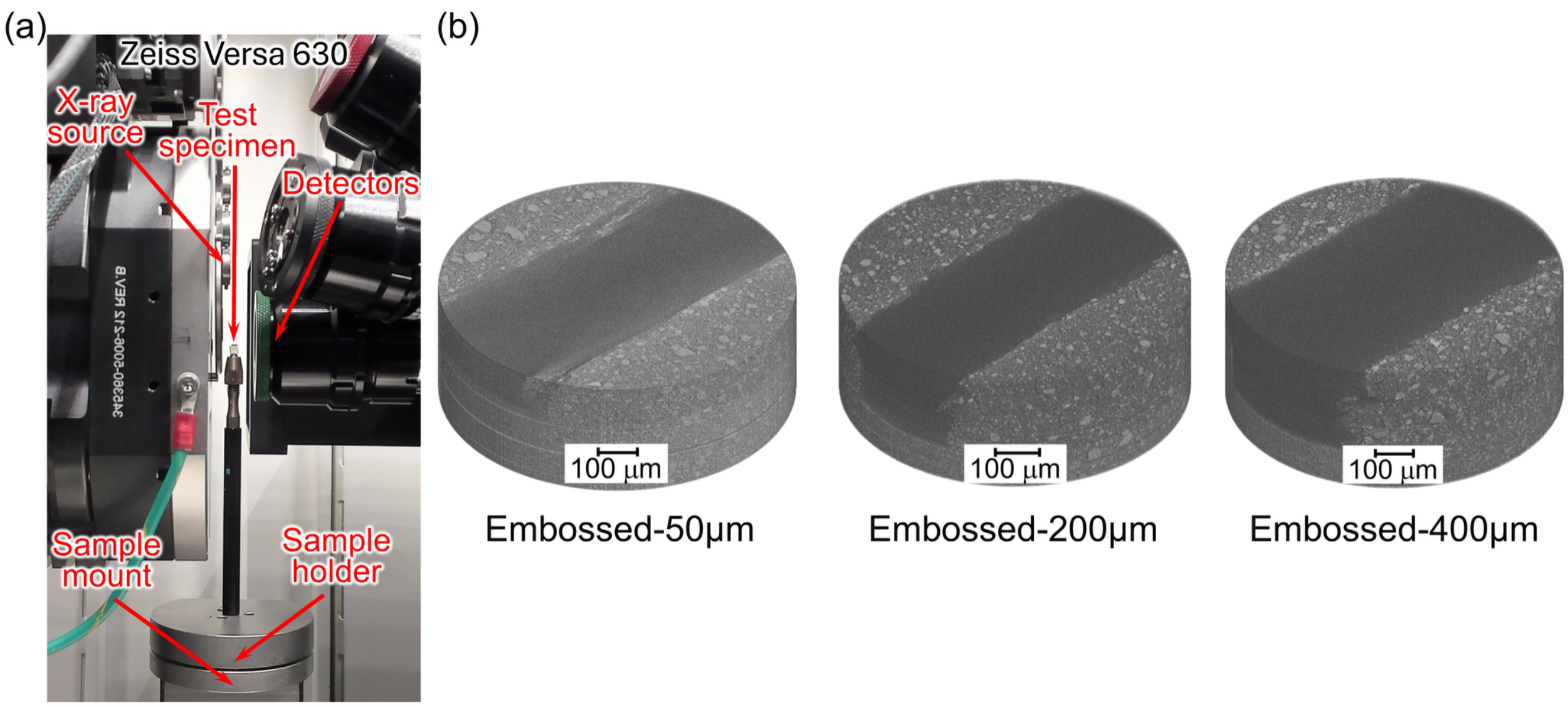

**Figure S8:** (a) X-ray CT setup in the Zeiss Versa 630 used for imaging embossed electrodes. (b) Representative rendered 3D reconstructions of Gen. III eµ-IDFF electrodes with embossing depths of 50, 200, and 400 µm. (scale bar = 100 µm)

**Table S3**: Scanning parameters with Zeiss Xradia Versa 630 scanner.

| Depth (µm) | Optical magnification | Binning | Exposure time (sec) | No. of projections (#) | Voltage (kV) | Power (W) | Source Filter | Voxel size (µm) | Scan time (h) |
|---|---|---|---|---|---|---|---|---|---|
| 50 | | | 4 | | | | LE1 | | 3.5 |
| 200 | 20 | 2 | 3 | 2001 | 80 | 10 | LE4 | 0.742 | 2.5 |
| 400 | | | | | | | LE2 | | |

## 8. Permeability Measurements of Patterned Electrodes

The hydraulic permeability of each patterned electrode was measured using two independent approaches: a gravity-driven apparatus and a pressure-driven apparatus. In the gravity-driven setup, a constant pressure head $p$ was maintained, as described in our previous work.[1,3] In the pressure-driven setup, a syringe pump (NE-300, NewEra Pump System Inc.) was used to impose controlled flow rates, and the resulting pressure difference was monitored (Fig. S9). In both cases, the apparent hydraulic permeability was determined from Darcy's law:[3,4]

$$k = \frac{\mu L_{e,\parallel} \dot{V}}{\Delta p A_c}$$

where $\mu$ is the dynamic viscosity of water (0.0890 mPa-s), $L_{e,\parallel}$ is the longitudinal length of the electrode, $A_c$ is the cross-sectional area of the electrode through which flow occurs, and $\dot{V}$ is the volumetric flow rate. The permeability values obtained from both methods are in close agreement, as summarized in Table S4, confirming the consistency and reliability of the measurements across different electrode types and fabrication methods.

**Table S4:** Electrode properties patterned with embossing and milling methods. The active material loading and porosity are calculated as described in our previous work.[3]

| Channel Gen. | Fabrication Method | PBA loading ($mg/cm^2$) | Porosity (%) | Permeability, $k_{app}$ ($\mu m^2$) Gravity-driven | Permeability, $k_{app}$ ($\mu m^2$) Pressure-driven |
|---|---|---|---|---|---|
| Gen. I | Milling[1] | 18.2 | 35.6 | 17 | - |
| | | 17.5 | 37.9 | 32 | - |
| Gen. II | Milling[1] | 19.1 | 32.6 | 70 | - |
| | | 20 | 28.8 | 65 | - |
| | Embossing | 16.32 | 38.26 | 22 | 23 |
| | | 15.2 | 42.5 | 25 | 29 |
| Gen. III | Embossing | 16.6 | 28 | 16 | 19 |
| | | 15.7 | 30.5 | 21 | 24 |
| | | 16.23 | 39.41 | 19 | 21 |
| | | 16.85 | 39.2 | 14 | 17 |

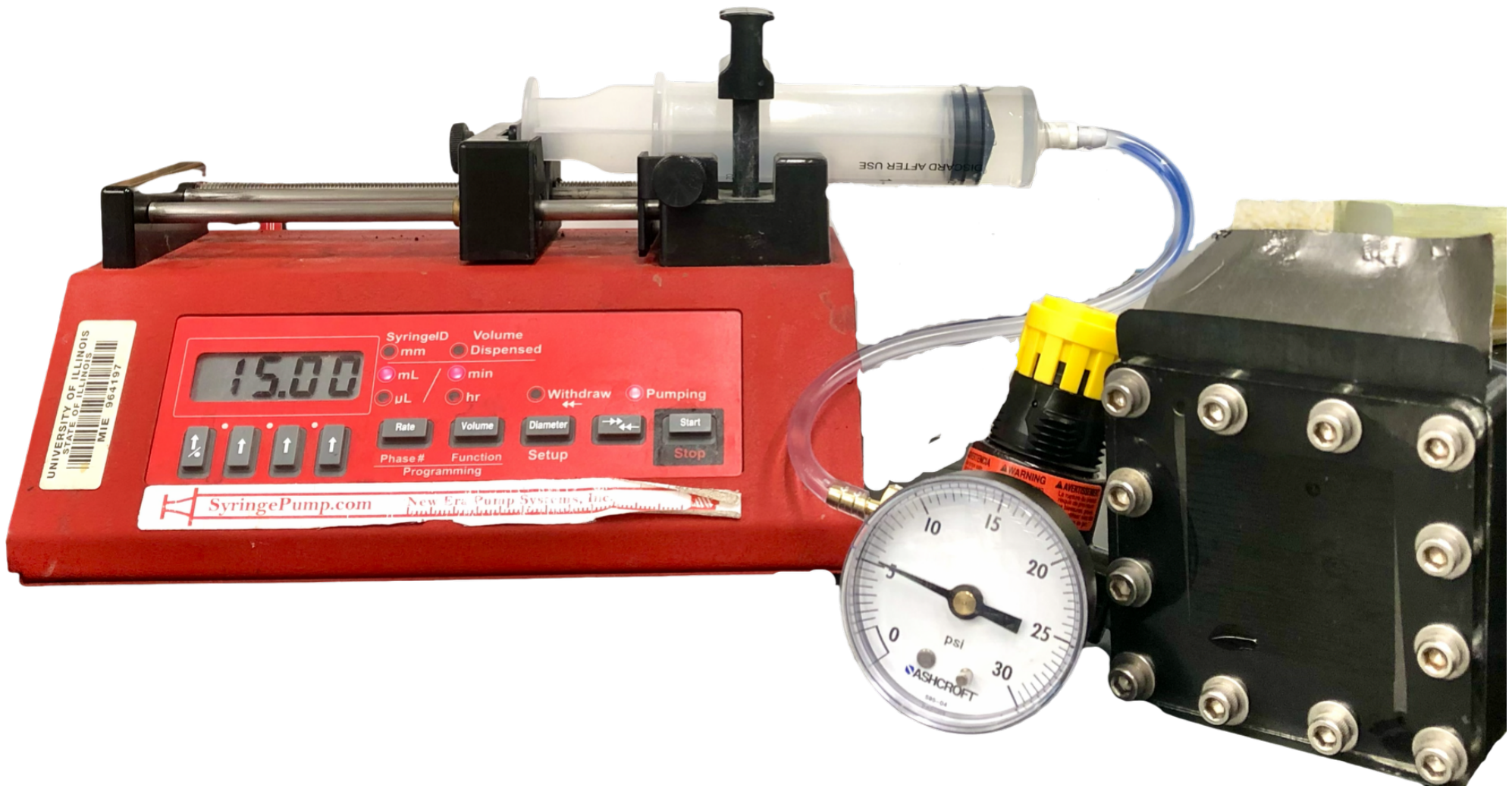

**Figure S9:** Pressure-driven setup to measure hydraulic permeability of electrodes patterned with both embossing and milling methods.

## 9. Rate Capability Tests for PBA Electrodes

To evaluate the electrochemical stability of the PBA electrodes patterned with Gen. II and Gen. III IDFFs, rate capability tests are conducted over a range of current densities between 0.1 and 8 mA/cm$^2$ at 1 and 5 mL/min flow rates, and compared with unpatterned electrodes.

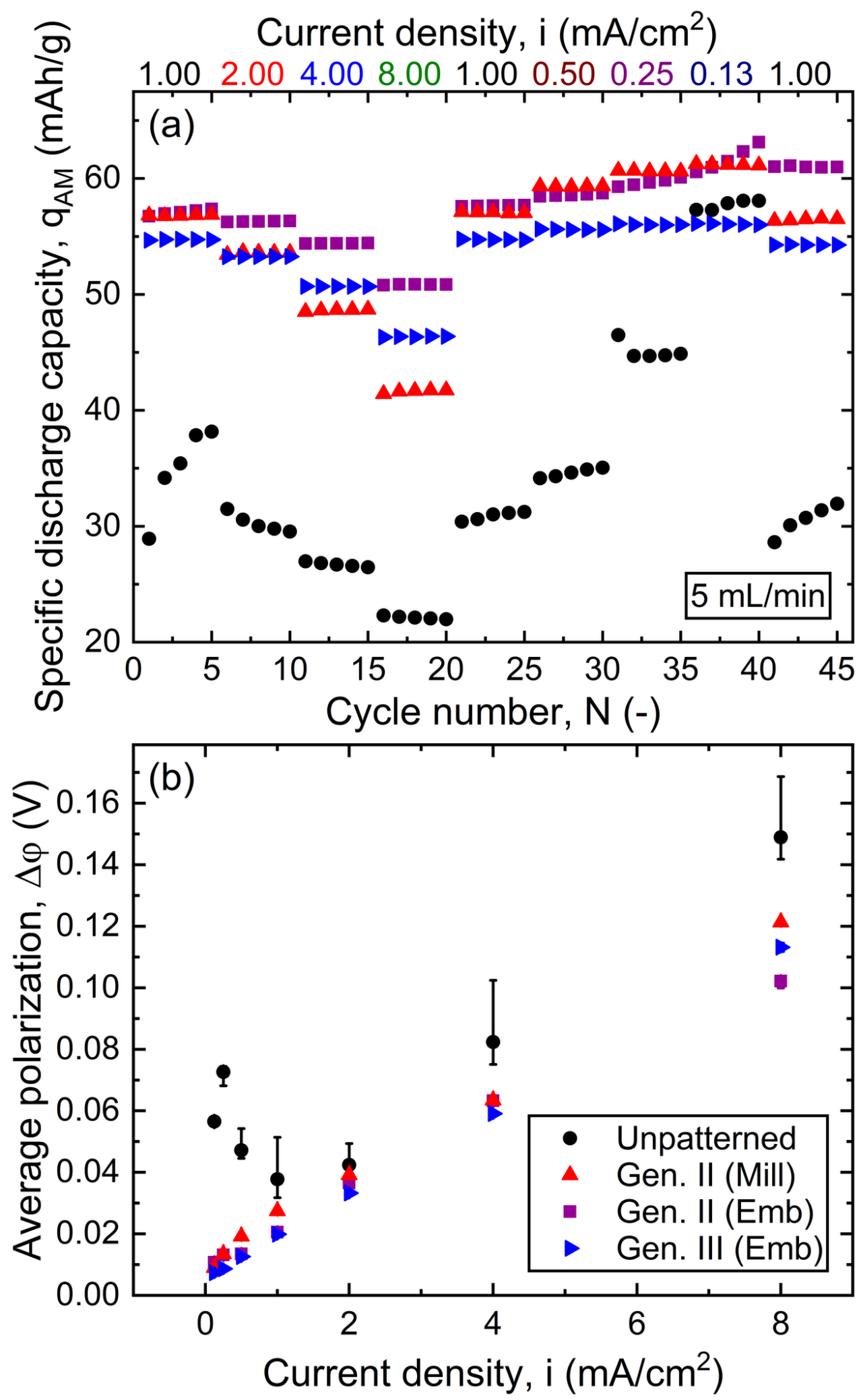

**Figure S10:** (a) Capacity retention and (b) average polarization during galvanostatic cycling at various rates for electrodes embossed with Gen. III eμ-IDFFs and either embossed or milled with Gen. II eμ-IDFFs. Unpatterned electrodes of the same thickness serve as a control. All electrodes used 5 mL/min flow rate of 500 mM NaCl solution drawn from and discharged to a common reservoir.

Figure S10 show the results at 5 mL/min flow rate, following trends similar to those observed at 1 mL/min (Fig. 3), demonstrating that the electrochemical benefits of the channel designs are maintained across different flow conditions.

## 10. Delay Time Optimization from Breakthrough Analysis

After each galvanostatic charge and discharge half-cycle, the inlet and outlet valves switch to direct the brine and diluate streams to the appropriate reservoirs. A delay time is introduced between the inlet and outlet valve switching events to minimize apparent intermixing between the two streams. If all valves are switched simultaneously, charge efficiency decreases because the brine and diluate remaining within the electrodes and connecting tubing can mix and flow into the wrong reservoir. In our previous studies, an initial estimate of the delay time was selected and the apparent mixing volume was measured at different flow rates.[1,3,5] The flow rate associated with the minimum mixing volume was then used to determine the optimized delay time for a given flow rate. However, this approach could yield multiple minima, making it difficult to identify a unique optimized delay time. To overcome this limitation, we adopted a more reliable method based on breakthrough analysis,[6] a transient flow characterization method in which concentration is monitored after a step change in inlet concentration, to determine the optimized delay time at different flow rates. The shape and onset of the breakthrough curve provide information about residence time, dispersion, mixing volume and flow distribution within the system.

For this breakthrough analysis, two reservoirs were used: a diluate reservoir containing deionized water $\left(V_{D_0} = 50\ mL\right)$ and a brine reservoir containing 500 mM NaCl $\left(V_{B_0} = 50\ mL\right)$. The effluent streams were delivered to their respective reservoirs over a range of flow rates from 1 to 10 mL/min, while the real-time salt concentration was monitred using four 3D-printed probes based on commercial micro-USB connectors connected to a Conduino apparatus positioned near the inlet and outlet ports.[7] Breakthrough experiments were conducted for electrodes embossed with Gen. II and Gen. III IDFFs. The corresponding normalized breakthrough curves are shown in Fig. S11a for both generations at 1 mL/min flow rate. From these results, we determined the median residence time $\tau_{50}$, defined as the time required for outlet concentration to reach 50% of its maximum value after the inlet valves are switched, and used this as the optimized delay time between valve-switching events (Fig. S11b).

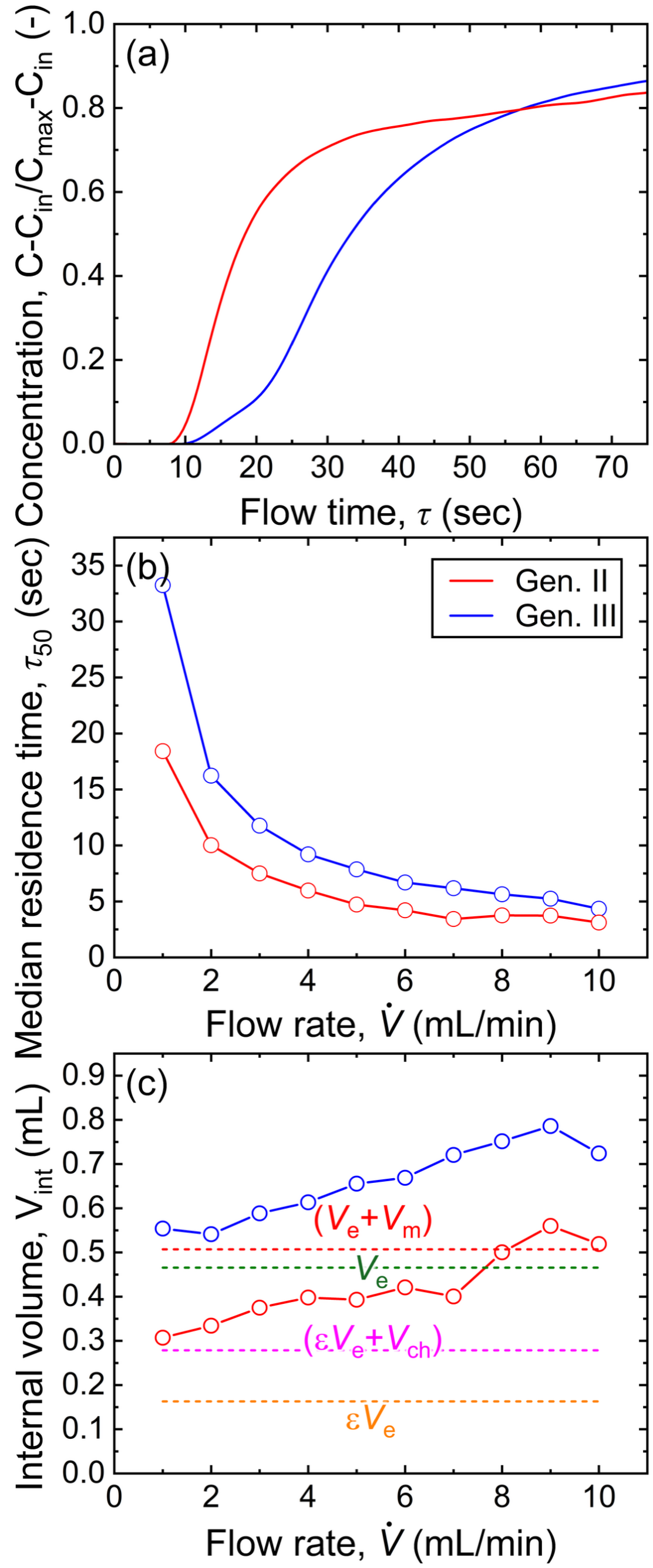

**Figure S11:** (a) Breakthrough curve showing the time variation of normalized outlet concentrations in diluate stream for electrodes embossed with Gen. II and Gen. III eμ-IDFFs at 1 mL/min flow rate. (b) Median residence time ($\tau_{50}$) as a function of flow rate calculated from the breakthrough analysis. (c) Internal volume of the FDI flow cell versus flow rate, with reference values corresponding to electrode volume ($V_e$), combined electrode and manifold volume ($V_e + V_m$), porosity-adjusted volumes ($\varepsilon V_e + V_{ch}$) and $\varepsilon V_e$, where $V_{ch}$ is the total channel volume and $\varepsilon$ is the electrode porosity.

At 1 mL/min flow rate, the optimized delay times were approximately 18.4 s and 33.0 s for the electrodes embossed with Gen. II and Gen. III eμ-IDFFs, respectively. We also calculated the internal volume of the effluents inside the system as the product of characteristic time and flow rate. This internal volume is then compared with the electrode volume, the porosity-adjusted electrode volume, the manifold volume, and the total channel volume (Fig. S11c). For Gen. III-embossed electrodes at 1 mL/min, the measured internal volume (0.55 mL) matches closely with the combined electrode and manifold volume (0.50 mL), with only 10% deviation. This agreement suggests that the optimized delay time is sufficient to remove nearly all of the solutoin from the system and direct it to the correct reservoir, thereby minimizing the apparent intermixing between effluents. In constrast, Gen. II-embossed electrodes showed greater apparent intermixing, consistent with its shorter optimized delay time. This behavior is likely associated with dead zones within the electrode resulting from the variable flow-path length between adjacent high- and low-pressure channels, which can cause the flow to preferentially follow shorter pathways and limit access to some regions of the electrode. These results further support the advantage of the Gen. III eμ-IDFF design in reducing dead zones and promoting more uniform flow distribution.

## 11. Water Recovery for Tapered-Channel eμ-IDFFs

Water recovery for seawater and brackish feeds is shown in Fig. S12 for Gen. II electrodes patterned by milling and embossing, and for Gen. III electrodes patterned by embossing alone.

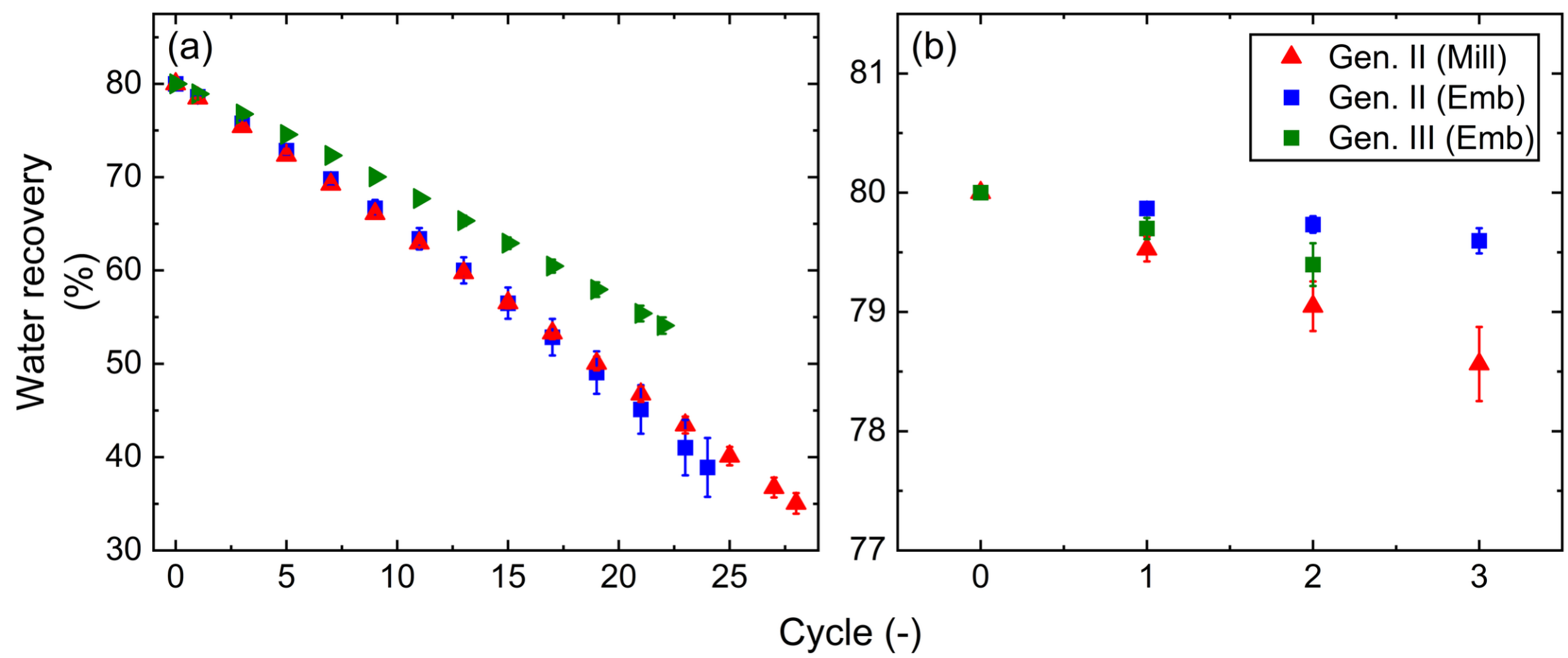

**Figure S12:** Comparison of water recovery for Gen. II eμ-IDFF electrodes patterned by milling and embossing, and for Gen. III eμ-IDFF electrodes patterned by embossing alone, with (a) seawater-salinity (SW, 29.2 g/L ≈ 500 mM NaCl) and (b) brackish-salinity (BW, 5.84 g/L ≈ 100 mM NaCl) feeds.

## References

(1) Rahman, M. H.; IV, I. C. L.; Do, V. Q.; Hamid, M. A.; Smith, K. C. Tapered, Interdigitated Channels for Uniform, Low-Pressure Flow Through Porous Electrodes for Desalination and Beyond. *Electrochim. Acta* **2025**, *514*, 145632. https://doi.org/10.1016/j.electacta.2024.145632

(2) Boussinesq. Mémoire Sur l'influence Des Frottements Dans Les Mouvements Réguliers Des Fluides. *J. Math. Pures Appl.* **1868**, *13*, 377–424.

(3) Do, V. Q.; Reale, E. R.; Loud, I. C.; Rozzi, P. G.; Tan, H.; Willis, D. A.; Smith, K. C. Embedded, Micro-Interdigitated Flow Fields in High Areal-Loading Intercalation Electrodes Towards Seawater Desalination and Beyond. *Energy Environ. Sci.* **2023**, *16* (7), 3025–3039. https://doi.org/10.1039/d3ee01302b

(4) Teng, Y.; Li, Z.; Chen, C. Review: Pre-Darcy Flows in Low-Permeability Porous Media. *Hydrogeology Journal 2024 32:8* **2024**, *32* (8), 1957–1977. https://doi.org/10.1007/s10040-024-02853-4

(5) Reale, E. R.; Regenwetter, L.; Agrawal, A.; Dardón, B.; Dicola, N.; Sanagala, S.; Smith, K. C. Low Porosity, High Areal-Capacity Prussian Blue Analogue Electrodes Enhance Salt Removal and Thermodynamic Efficiency in Symmetric Faradaic Deionization with Automated Fluid Control. *Water Res. X* **2021**, *13*. https://doi.org/10.1016/j.wroa.2021.100116

(6) Marsh, H.; Rodríguez-Reinoso, F. Characterization of Activated Carbon. *Activated Carbon* **2006**, 143–242. https://doi.org/10.1016/b978-008044463-5/50018-2

(7) Carminati, M.; Luzzatto-Fegiz, P. Conduino: Affordable and High-Resolution Multichannel Water Conductivity Sensor Using Micro USB Connectors. *Sens. Actuators B Chem.* **2017**, *251*, 1034–1041. https://doi.org/10.1016/j.snb.2017.05.184